\documentclass[reprint,amsmath,amssymb,aps,prb,groupedaddress,nofootinbib,twocolumn,superscriptaddress]{revtex4-2}
\usepackage{graphicx,xcolor}
\usepackage{dcolumn}
\usepackage{bm}
\usepackage{paralist}
\usepackage{hhline}
\usepackage{gensymb}
\usepackage{hyperref}
\usepackage{tikz, rotating, array, longtable, adjustbox, booktabs} 
\usepackage{comment}

\begin{document}

\title{Fermi surface and magnetic breakdown in PdGa}
\author{Nico Huber}
\email{nico.huber@tum.de}
\affiliation{Technical University of Munich, TUM School of Natural Sciences, Physics Department, D-85748 Garching, Germany}
\author{Ivan Volkau}
\affiliation{Technical University of Munich, TUM School of Natural Sciences, Physics Department, D-85748 Garching, Germany}
\author{Alexander Engelhardt}
\affiliation{Technical University of Munich, TUM School of Natural Sciences, Physics Department, D-85748 Garching, Germany}
\author{Ilya Sheikin}
\affiliation{Laboratoire National des Champs Magn\'{e}tiques Intenses (LNCMI-EMFL), CNRS, Univ. Grenoble Alpes, 38042 Grenoble, France}
\author{Andreas Bauer}
\affiliation{Technical University of Munich, TUM School of Natural Sciences, Physics Department, D-85748 Garching, Germany}
\affiliation{Technical University of Munich, TUM Center for Quantum Engineering (ZQE), D-85748 Garching, Germany}
\author{Christian Pfleiderer}
\affiliation{Technical University of Munich, TUM School of Natural Sciences, Physics Department, D-85748 Garching, Germany}
\affiliation{Technical University of Munich, TUM Center for Quantum Engineering (ZQE), D-85748 Garching, Germany}
\affiliation{Munich Center for Quantum Science and Technology (MCQST), D-80799 Munich, Germany}
\author{Marc A. Wilde}
\email{marc.wilde@tum.de}
\affiliation{Technical University of Munich, TUM School of Natural Sciences, Physics Department, D-85748 Garching, Germany}
\affiliation{Technical University of Munich, TUM Center for Quantum Engineering (ZQE), D-85748 Garching, Germany}

\date{\today}

\begin{abstract}
We study the electronic structure of the chiral semimetal PdGa by means of the de~Haas-van~Alphen and Shubnikov-de~Haas effect. We find that the Fermi surface of PdGa comprises multiple pockets split by spin-orbit coupling. We compare our experimental findings with the band structure calculated \emph{ab initio}. We demonstrate that the quantum oscillation spectra can be fully understood by considering nodal plane degeneracies at the Brillouin zone boundary and magnetic breakdown between individual Fermi surface pockets. Expanding traditional analysis methods, we explicitly calculate magnetic breakdown frequencies and cyclotron masses while taking into account that extremal breakdown trajectories may reside away from the planes of the single-band orbits. We further analyze high-frequency contributions arising from breakdown trajectories involving multiple revolutions around the Fermi surface which are distinct from conventional harmonic frequencies. Our results highlight the existence of gaps induced by spin-orbit coupling throughout the band structure of PdGa, the relevance of nodal planes on the Brillouin zone boundary, and the necessity for a comprehensive analysis of magnetic breakdown.
\end{abstract}

\maketitle

\section{Introduction}
Chiral compounds with cubic crystal structure exhibit interesting optical \cite{2020_Rees_SciAdv, 2023_Hsu_PhysRevB, 2024_Brinkman_PhysRevLett} and transport properties \cite{2023_Yang_ProcNatlAcadSci,2024_Balduini_NatCommun}. These properties are directly linked to the topology of the electronic band structures that feature characteristic spin-momentum textures \cite{2018_Chang_NatureMater}.
Under the influence of spin-orbit coupling, the lack of inversion symmetry leads to a splitting of bands at generic points in reciprocal space. This splitting changes the nature of topological crossings \cite{2017_Tang_PhysRevLett} and consequently the physical responses of the system.
B20 compounds, crystallizing in space group $P2_13$ (198), exhibit a wide range of electronic and magnetic properties and comprise many examples with interesting characteristics \cite{2013_Nagaosa_NatureNanotech, 2017_Chang_PhysRevLett, 2017_Chang_PhysRevLett, 2018_Fang_ProcNatlAcadSci, 2019_Klotz_PhysRevB, 2019_Schroter_NatPhys, 2019_Rao_Nature, 2021_Ni_NatCommun, 2021_Ma_NatCommuna, 2023_Huber_Nature}.
As B20 compounds form from various light and heavy transition metal constituents, they exhibit electronic structures with varying degrees of spin-orbit coupling-induced band splitting. Understanding these electronic structures and their geometrical properties is crucial when searching for materials relevant for spin-related applications.

PdGa is a non-magnetic, semimetallic member of the B20 family \cite{2012_Klanjsek_JPhysCondensMatter}. Angle-resolved photoemission spectroscopy confirmed the coarse band structure predicted by first-principles calculations and inferred a maximal topological charge of $|C|=4$ for the multifold band crossings located at the $\Gamma$- and the R-point \cite{2020_Schroter_Science}.
Circular dichroism measurements combined with band structure calculations revealed 
a polar orbital momentum texture around the R-point \cite{2024_Yen_NatPhys} promising a large orbital Hall and orbital magnetoelectric effect \cite{2023_Yang_ProcNatlAcadSci}.
Quantum oscillation studies probing the de Haas-van Alphen effect reported on a multitude of frequency contributions, reflecting the complex Fermi surface of PdGa \cite{2022_Zeng_PhysRevB, 2025_Miyake_JPhysSocJpn}.

The crystalline symmetries of space group 198 enforce extended topological band degeneracies on the Brillouin zone boundary without an upper limit for their topological charge \cite{2021_Wilde_Nature, 2022_Huber_PhysRevLett}. These degeneracies, termed nodal planes, distinctly influence the Fermi surface geometry and are crucial in the interpretation of quantum oscillation spectra. In both previously reported quantum oscillation studies on PdGa \cite{2022_Zeng_PhysRevB, 2025_Miyake_JPhysSocJpn} the influence of nodal planes was not taken into account.

Due to gaps between individual Fermi surface pockets induced by spin-orbit coupling, quasiparticle trajectories in many B20 systems are additionally prone to magnetic breakdown. Because of the low symmetry of space group 198, extremal breakdown trajectories may be located outside planes containing high-symmetry points or planes on which the constituent orbits are extremal. While the search for extremal trajectories away from high-symmetry planes is implemented in widely utilized tools such as SKEAF \cite{2012_Rourke_ComputerPhysicsCommunications}, there is no equivalent code for magnetic breakdown trajectories. A comprehensive interpretation of quantum oscillation spectra in low-symmetry materials with small gaps requires an algorithm searching for these extremal breakdown trajectories.
We note that magnetic breakdown at the R-point of PdGa was analyzed in Ref.~\cite{2025_Miyake_JPhysSocJpn}, albeit only on a qualitative level.

Complete magnetic breakdown between pockets centered around the R-point was observed in the light transition metal silicide CoSi. The underlying small gap has been discussed in the context of quasi-symmetries \cite{2022_Guo_NatPhys}. This raises the question of whether other B20 compounds feature similarly small gaps stabilized by quasi-symmetries and highlights the necessity for a comprehensive analysis of magnetic breakdown phenomena in this material family.

Here, we present a study of de~Haas-van~Alphen and Shubnikov-de~Haas oscillations in PdGa. We compare our experimental results with \emph{ab initio} band structure calculations that take into account the nodal plane degeneracies on the Brillouin zone boundary. We developed a code to calculate magnetic breakdown frequencies for extremal trajectories that may reside outside of the planes containing the individual extremal single-band orbits. We find an excellent match between the experimentally observed and calculated frequencies arising from Fermi surface pockets around the R-, M-, and $\Gamma$-point. We further analyzed high-frequency contributions originating in magnetic breakdown orbits that involve multiple revolutions around the Fermi surface at the R-point. Our results show that spin-orbit coupling-induced gaps exist throughout the Brillouin zone of PdGa and demonstrate that the quantum oscillation spectra can be fully understood by considering the influence of nodal planes and a comprehensive analysis of magnetic breakdown trajectories.

Our paper is organized as follows: Sec.~\ref{sec:ElectronicStructure} presents the electronic band structure of PdGa as calculated from first principles. An overview of the experimental methods and data analysis is reported in Sec.~\ref{sec:ExperimentalMethods}. The experimental results, including a comparison of the quantum oscillations and the Fermi surface of PdGa, are presented in Sec.~\ref{sec:ExperimentalResults}. A detailed analysis of magnetic breakdown frequencies is given in Sec.~\ref{sec:Discussion}. The paper closes with a concluding discussion in Sec.~\ref{sec:Conclusion}

\section{Electronic structure}\label{sec:ElectronicStructure}

\begin{figure*}
    \begin{center}
        \includegraphics[width=\textwidth]{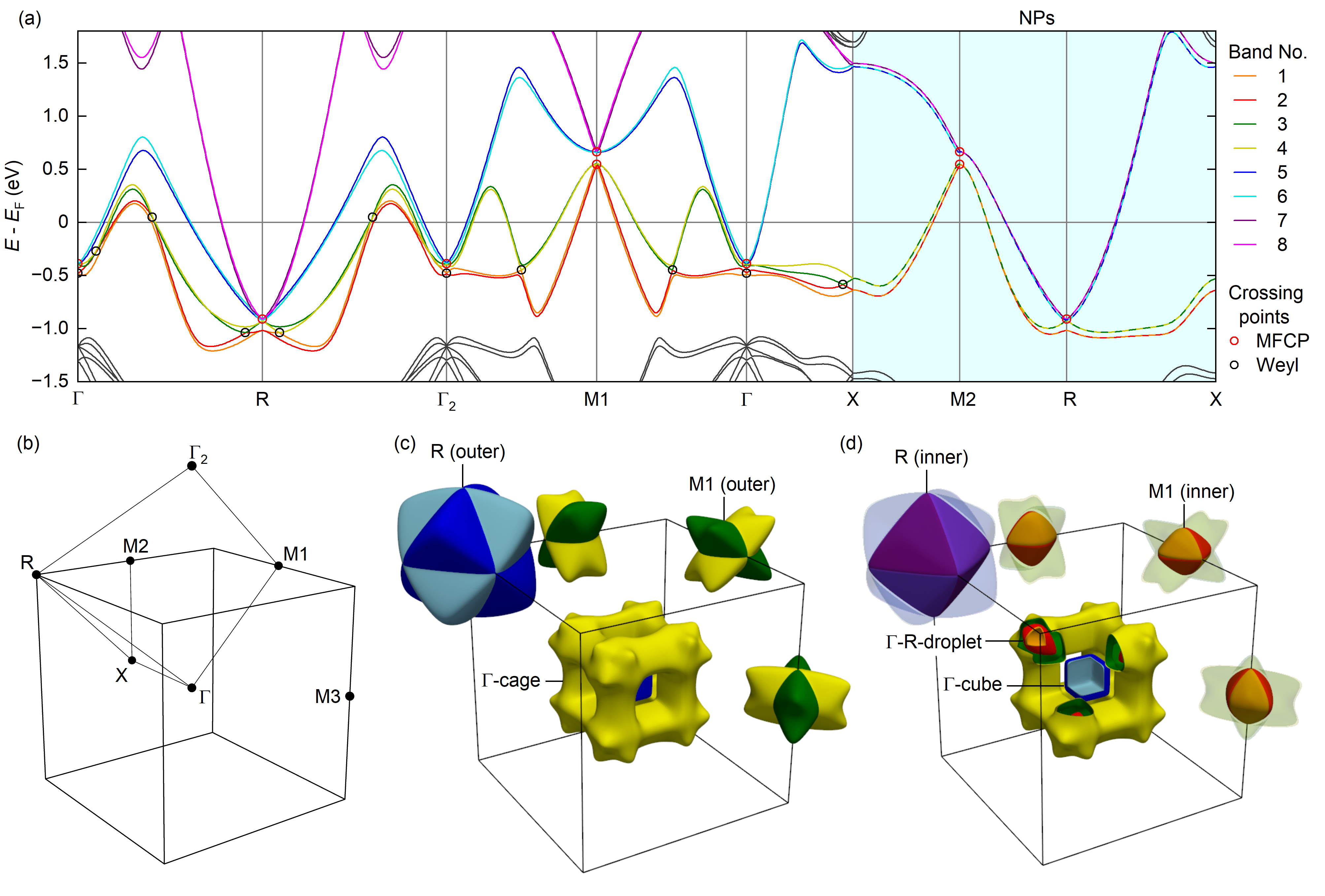}
        \caption{Electronic structure of PdGa. (a)~Ab initio calculated band structure along the path given in (b). Spin-orbit coupling was included. Eight bands cross the Fermi level and are interconnected by multifold crossing points (MFCP) at $\Gamma$, M, and R, Weyl points at high-symmetry lines, and nodal planes (NPs) on the Brillouin zone~(BZ) boundary (shaded in cyan). (b)~Path used for the band structure calculation shown in (a) with high-symmetry points indicated. The $\Gamma_2$-point in the second BZ is included to highlight the role of NPs, namely an inversion of bands in neighboring BZs. We denote the three distinct M-points M1-M3 to facilitate the discussion of the different orientations of the respective Fermi surface (FS) pockets. (c)~FS of PdGa comprising multiple nested pockets around the high-symmetry points $\Gamma$, R, and M. (d)~FS with outer pockets drawn semi-transparent and selective cutouts to permit a view on the inner FS pockets.
      }
      \label{fig:bands-FS}
  \end{center}
\end{figure*}

PdGa crystallizes in the chiral, simple cubic Sohncke space group $P2_13$ (198). Both Pd and Ga occupy Wyckoff positions $4a$ at coordinates $(u,u,u)$, $(-u+1/2,-u,u+1/2)$, $(-u,u+1/2,-u+1/2)$ and $(u+1/2,-u+1/2,-u)$ with $u_{\mathrm{Pd}}=0.142$ and $u_{\mathrm{Ga}}=0.843$.
We calculated the electronic band structure with density functional theory (DFT) using WIEN2k \cite{WIEN2k} and the generalized gradient approximation \cite{1996_Perdew_PhysRevLett} taking into account spin-orbit coupling (SOC). The calculations were converged on a 22x22x22 Monckhorst-Pack grid. For the calculation of the Fermi surface (FS), band energies were sampled on a grid of 100x100x100 $k$-points in the full Brillouin zone~(BZ). The lattice constant was fixed to the experimentally observed value $a=4.896$\,\AA \cite{2020_Schroter_Science}.

The calculated band structure is shown in Fig.~\ref{fig:bands-FS}(a). In total, eight bands cross the Fermi level as depicted in eight different colors. Due to the lack of inversion symmetry, SOC leads to a band splitting of up to 100\,meV at arbitrary points in the Brillouin zone. The crystalline symmetries enforce degeneracies at high-symmetry points and planes, thus connecting the bands at multifold topological crossings located at the $\Gamma$-, M-, and R-points and topological nodal planes (NPs) at the Brillouin zone (BZ) boundary \cite{2016_Bradlyn_Science, 2022_Huber_PhysRevLett}. Further, twofold band crossings in the form of Weyl points exist along the high-symmetry lines, e.g., between bands 2 and 3 on the $\Gamma$-R line. The multifold crossing points (MFCPs) and selected Weyl points are indicated in Fig.~\ref{fig:bands-FS}(a) by circles. A full analysis of topological crossings is beyond the scope of this paper.
The pairwise degeneracy enforced by NPs is visible along the X-M2-R-X path, where degenerate bands are drawn as dashed lines. To further illustrate the effect of the NPs on the band structure, we include the $\Gamma_2$-point of the second BZ in the path shown in Figs.~\ref{fig:bands-FS}(a) and \ref{fig:bands-FS}(b). Because the bands intersect at the NPs, they switch in energy upon crossing the BZ boundary. This effect strongly influences the shape of the FS pockets centered around the M- and R-point and consequently the extremal trajectories probed by quantum oscillations \cite{2021_Wilde_Nature, 2022_Huber_PhysRevLett, 2024_Huber_PhysRevB}.

The FS of PdGa is shown in Figs.~\ref{fig:bands-FS}(c) and \ref{fig:bands-FS}(d). In Fig.~\ref{fig:bands-FS}(d), the outer FS pockets are drawn semi-transparent and selective cutouts are excluded to allow for a view of the inner pockets.
The FS around the $\Gamma$-point consists of two cuboid-shaped electron pockets in the center and two cage-like hole pockets located around them.
Nested inside the corners of the cage structure are eight pairs of droplet-shaped hole pockets. The outer droplet almost touches the inner cage pocket due to a Weyl point located extremely close to the Fermi level along the $\Gamma$-R line.

There are two pairs of hole-like pockets centered around each M-point. Within each pair, the FS pockets are separated by a SOC-induced gap but intersect each other due to the symmetry-enforced NPs on the BZ boundary. Three symmetry-related copies of the pockets centered around the M-points exhibit different orientations with respect to an externally applied magnetic field.

The FS around the R-point also consists of two pairs of nested pockets with the pockets of a pair mutually intersecting each other at the BZ boundary. They are electron-like and larger than the pockets centered around the M-point. The inner pair of pockets is octahedral-shaped and exhibits a small SOC-induced splitting. The outer pair of pockets features `lobes' along the [111] directions and shows a slightly larger splitting.

\section{Experimental methods}\label{sec:ExperimentalMethods}

\subsection{Sample synthesis}
Polycrystalline rods of PdGa with a diameter of 6\,mm were synthesized from a stoichiometric mixture of Pd (4N5) and Ga (6N) in an inductively heated rod casting furnace \cite{2016_Bauer_RevSciInstrum}. The composition was confirmed via powder X-ray diffraction. A single-crystalline ingot of PdGa was grown from the poly-crystalline rods using an optical floating-zone furnace under 5\,bar Ar atmosphere with a constant flow of 0.1\,L\,min$^{-1}$ to reduce evaporation from the molten zone. The feed and seed rod were rotated in opposite directions at 10\,rpm and the molten zone moved through the crystal at a rate of 5\,mm\,h$^{-1}$.
The resulting single crystal was oriented using Laue X-ray diffraction and cut into several platelets with axes along high-symmetry directions and approximate dimensions $4 \times 1 \times 0.2$\,mm$^3$. To further improve the crystalline quality, the platelets were subsequently annealed for 100\,h at 900\degree C under 1\,bar Ar atmosphere. The sample exhibiting the largest residual resistivity ratio (RRR) of $\approx 100$ was used for all experiments in magnetic fields up to 18\,T presented in this work. High-field torque magnetometry experiments were performed on a different sample from the same growth prior to annealing.

\subsection{Detection of quantum oscillations}
Shubnikov–de Haas (SdH) oscillations in the transverse magnetoresistance were recorded in magnetic fields up to 18\,T.
Measurements were performed at constant temperatures between 1.5 to 30\,K in a Helium flow cryostat with the magnetic field oriented along the crystallographic [110] direction. Further measurements down to 0.1\,K with the same field orientation were performed in a dilution refrigerator.
The angular dependence of the SdH oscillations was investigated at a constant temperature of 1.5\,K. The sample was rotated around the [1$\bar{1}$0] direction in steps of $4^\circ$ over a range of $90^\circ$, resulting in a rotation of the magnetic field direction between [001] and [110].
Electrical contacts to the sample were established using Al wire bonds. An AC excitation current with a frequency of 22.08\,Hz and an amplitude of 4\,mA\,rms was applied. The longitudinal voltage drop across the sample was recorded using low-noise transformers and a conventional lock-in technique.

Cantilever-based torque magnetometry with capacitive readout was used to measure the de~Haas–van~Alphen (dHvA) effect in magnetic fields up to 18\,T. The temperature dependence of the dHvA oscillation amplitudes between 1.5 and 30\,K was measured with the magnetic field direction slightly tilted away from the [110]-direction in order to provide a finite torque. The angular dependence of dHvA oscillations was again recorded in the (1$\bar{1}$0) rotation plane.
Additionally, high-field torque magnetometry experiments were conducted at the LNCMI Grenoble in magnetic fields up to 36\,T at $T<0.1$\,K.

We note that torque magnetometry is sensitive to the magnetization perpendicular to an applied magnetic field and the oscillation amplitudes scale with the torque reduction factor
\begin{equation}
    R_\mathrm{Torque} = \frac{1}{f}\frac{df}{d\theta} \quad ,
    \label{eq:Torque-factor}
\end{equation}
where $f$ is the dHvA frequency \cite{1984_Shoenberg_Book}.
For all measurements the angle $\theta$ is defined as the angle between the magnetic field orientation and the [001] direction in a (1$\bar{1}$0) rotation plane.

\subsection{Data analysis}
Both SdH and dHvA measurements were analyzed as follows. First, the oscillatory part of the signal was obtained by subtracting a 4th-order polynomial fit from the raw data to remove the non-oscillatory signal contribution. The oscillatory part of the signal was then interpolated onto evenly spaced values of the inverse applied magnetic field, multiplied by a Hamming window to suppress spectral leakage, and zero-padded to increase the Fourier sampling rate. Peaks in the Fast Fourier Transforms (FFT) were fitted with Gaussians to obtain their amplitudes and frequencies.

The temperature dependence of quantum oscillation amplitudes was analyzed within the Lifshitz-Kosevich (LK) formalism. The temperature reduction factor $R_T$ is given by
\begin{equation}
    R_T = \frac{X}{\sinh{X}} \hspace{0.5cm} \mathrm{with} \hspace{0.5cm} X = \frac{2\pi^2 p m^* k_B T }{\hbar e B} \quad ,  
    \label{eq:LK-factor}
\end{equation}
where $m^*$ is the cyclotron mass, $k_B$ the Boltzmann constant, $T$ the temperature, $e$ the electron charge,  $p$ the harmonic number, and $\frac{1}{B}$ the mean of the inverse applied magnetic field in the analyzed field range \cite{1984_Shoenberg_Book}.

\subsection{Calculation of QO frequencies}\label{sec:Methods_MB}
\begin{figure}
  \begin{center}
  \includegraphics[width=\columnwidth]{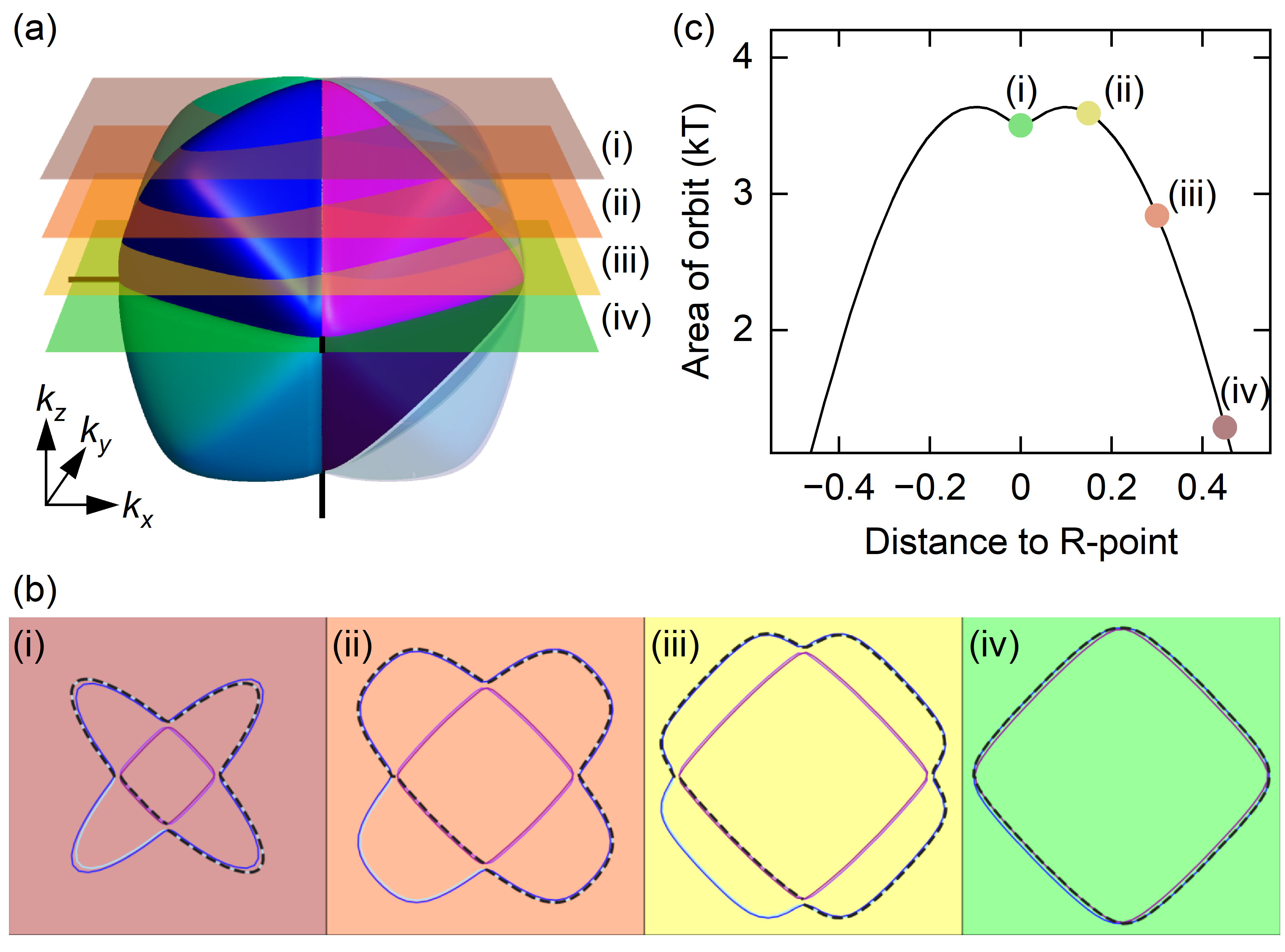}
  \caption{
  Procedure used for the calculation of magnetic breakdown~(MB) branches. (a)~FS of PdGa at the R-point comprising four pockets with exemplary slices (i)-(iv) at which the FS cross sections were calculated. The right side of the outer FS pockets is drawn semi-transparent to permit a view of the inner pockets. (b)~FS cross sections on the slices shown in (a). An exemplary MB orbit is shown as a dashed line. (c)~Caluclated areas of the MB orbit as a function of distance to the R-point. Colored dots correspond to the slices. With the field applied along [001] two frequencies are expected from the minimum in the plane crossing the R-point and the two degenerate maxima above and below.
  }
  \label{fig:Extremal_MB_orbits_calculation}
  \end{center}
  \end{figure}

To compare the experimentally detected quantum oscillations to our DFT results we extracted the frequencies and cyclotron masses from the band energies calculated on a dense $k$-grid using the SKEAF tool \cite{2012_Rourke_ComputerPhysicsCommunications}. For the calculation of magnetic breakdown trajectories, we developed a Python script performing the procedure outlined in the following. Figure~\ref{fig:Extremal_MB_orbits_calculation} illustrates the steps involved on the example of the FS pockets of PdGa centered around the R-point.

First, the FS was sliced at equidistant planes parallel to the fixed direction of the magnetic field, cf. Fig.~\ref{fig:Extremal_MB_orbits_calculation}(a) with exemplary slices denoted (i) through (iv). Second, the positions of magnetic breakdown~(MB) junctions were calculated for each cross section by searching for minima in the distance between the trajectories. Third, a particular MB was selected [dashed line in Fig.\ref{fig:Extremal_MB_orbits_calculation}(b)] and its area calculated as a function of $k_\parallel$ along the magnetic field direction. Analogously to fundamental orbits, the extremal cross sections of these MB orbits correspond to QO frequencies.
Fourth, once an extremal cross section was found, the curvature of the MB trajectory at the extremal cross section was calculated from its $k_\parallel$-dependence and the effective mass was extracted by varying the chemical potential around $E_\mathrm{F}$.
Fifth, the breakdown probabilities $P_i$ at each breakdown junction were evaluated using Chamber's formula \cite{1966_Chambers_ProcPhysSoc}
\begin{equation}
    P_i=e^{-\frac{B_0}{B}} \quad \mathrm{with} \quad B_0=\frac{\pi \hbar}{2 e}\sqrt{\frac{k_g^3}{a+b}} \quad ,
    \label{eq:Chambers}
\end{equation}
where $k_g$ is the gap in $k$-space and $a$ and $b$ are the curvatures of the orbits at the breakdown junction. The total probability for a particular orbit is given as the product of the probabilities at each junction.

The procedure was repeated for all possible MB orbits while varying the angle of the applied magnetic field.
To permit an easy comparison with the experimental data, colormaps of the MB branches were calculated by weighing the extracted frequencies with an amplitude prefactor given by
\begin{equation}
    A=\frac{1}{m^* \sqrt{c}} \prod_i P_i \quad .
\end{equation}

We note that we calculated the MB trajectories at the M- and R-point between one inner and one outer trajectory (instead of the actual four). This simplification is permitted due to the degeneracies associated with the nodal planes at the BZ boundary. These degeneracies relate trajectories pairwise by symmetry. Hence, the calculated frequency spectrum is equivalent for different combinations of inner and outer trajectories. Considering MB between all four trajectories may, however, change the degeneracy of MB frequencies and thus influence the observed QO amplitudes.

\section{Experimental Results}\label{sec:ExperimentalResults}

In this section, we present the results of our Shubnikov-de~Haas~(SdH) and de~Haas-van~Alphen~(dHvA) studies and connect the observed quantum oscillations to extremal trajectories on the FS of PdGa. In this section, we focus on quantum oscillation frequencies that may be explained by extremal trajectories on a single band without magnetic breakdown. A detailed discussion of quantum oscillations arising from magnetic breakdown trajectories is presented in Sec.~\ref{sec:Discussion}.

\subsection{Magnetoresistance}

\begin{figure}
  \includegraphics[width=\linewidth]{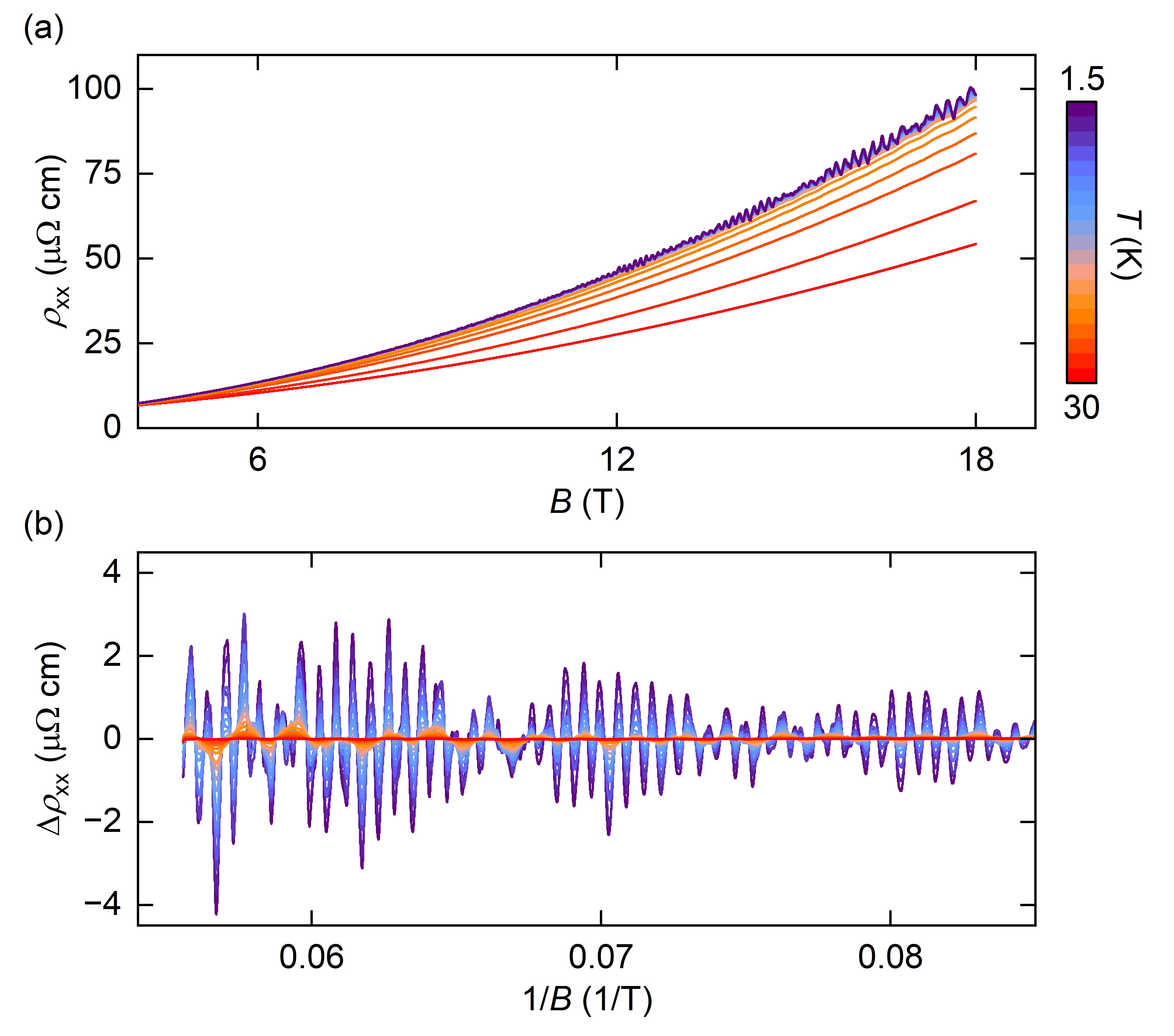}
  \caption{Typical magnetoresistance of PdGa. (a)~Transverse magnetoresistance, $\rho_{xx}$, as a function of magnetic field $B\parallel [110]$ at different temperatures between 1.5 and 30\,K. (b)~Oscillatory component of the magnetoresistance, $\Delta \rho_{xx}$, as a function of the inverse applied magnetic field, $1/B$.
  }
  \label{fig:SdH_high_T_overview}
\end{figure}

Figure~\ref{fig:SdH_high_T_overview}(a) shows the resistivity, $\rho_{xx}$, of PdGa in the transverse magnetoresistance~(MR) configuration ($B \parallel [110], I\parallel [1\bar{1}0]$) recorded at different temperatures between 1.5\,K and 30\,K. $\rho_{xx}$ increases nearly quadratically up to 18\,T, as expected for a compensated semimetal. The magnitude of the magnetoresistance decreases with increasing temperature. At low temperatures and high magnetic fields, characteristic SdH oscillations emerge. The oscillatory part of the resistivity, $\Delta\rho_{xx}$, after subtraction of a polynomial is shown in Fig.~\ref{fig:SdH_high_T_overview}(b). The oscillations are periodic in $1/B$ and exhibit multiple frequencies. The oscillation amplitudes decrease with increasing temperature.

\subsection{Torque magnetometry}
\begin{figure}
  \includegraphics[width=\linewidth]{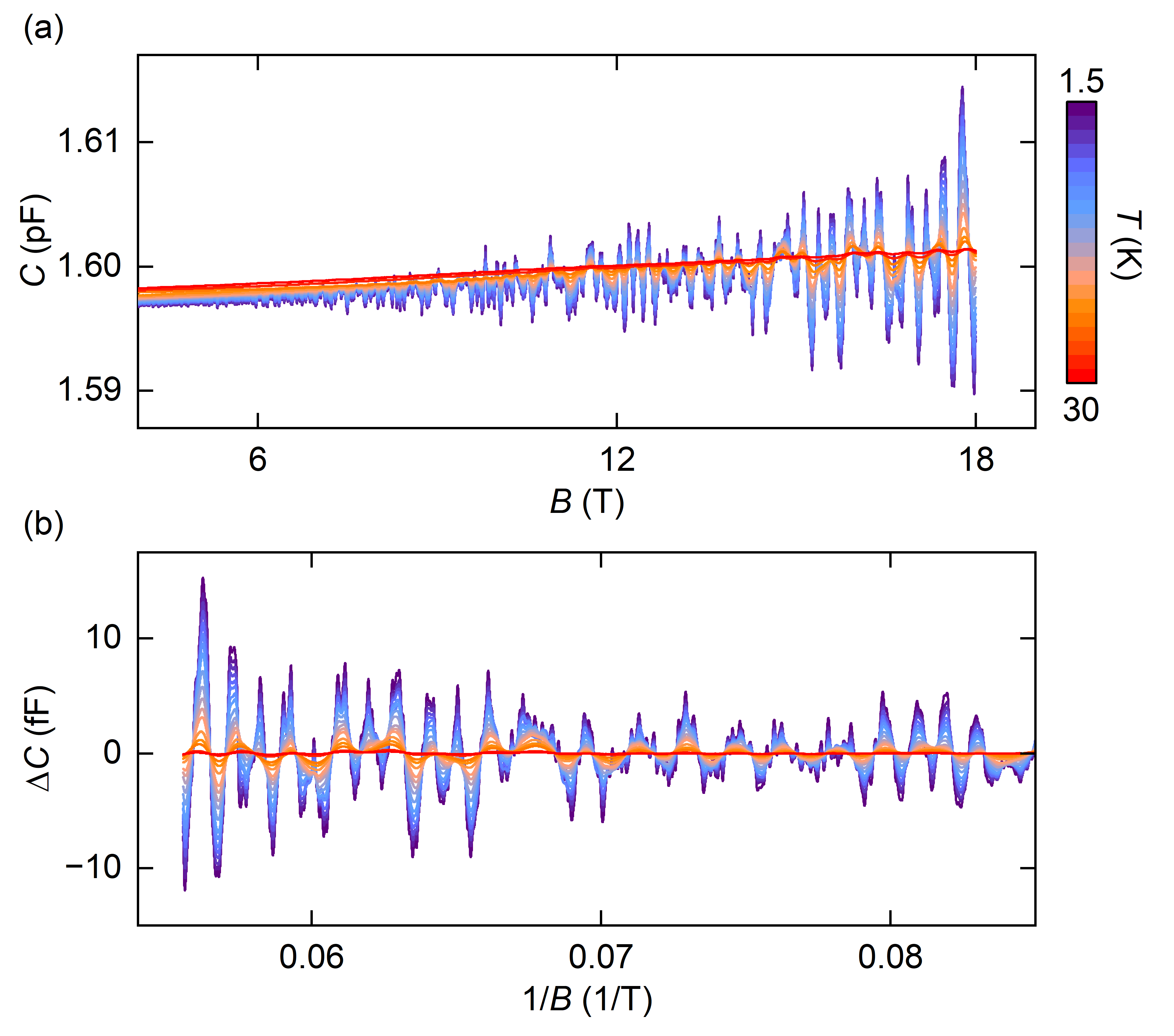}
  \caption{Typical torque magnetometry data recorded at TUM. (a)~Capacitance, $C$, of the torque magnetometer as a function of magnetic field at $\theta = 86^\circ$ at different temperatures between 1.5 and 30\,K. (b)~Oscillatory component of the capacitance, $\Delta C$, as a function of inverse applied magnetic field, $1/B$.
  }
  \label{fig:dHvA_high_T_overview}
\end{figure}

The magnetization of PdGa was measured by means of cantilever-based torque magnetometry using a capacitive readout. The capacitance, $C$, as a function of applied magnetic field is shown in Fig.~\ref{fig:dHvA_high_T_overview}(a). Data at different temperatures between 1.5 and 30\,K was recorded at $\theta=86^\circ$, corresponding to $B$ applied close to the $[110]$ direction. The signal exhibits strong dHvA oscillations on a monotonically increasing, almost temperature independent background. The oscillatory part of the capacitance, $\Delta C$, as a function of inverse applied magnetic field is depicted in Fig.~\ref{fig:dHvA_high_T_overview}(b). Again, a complex pattern of $1/B$-periodic oscillations is observed whose amplitudes decrease with increasing temperature.

\begin{figure}
  \includegraphics[width=\linewidth]{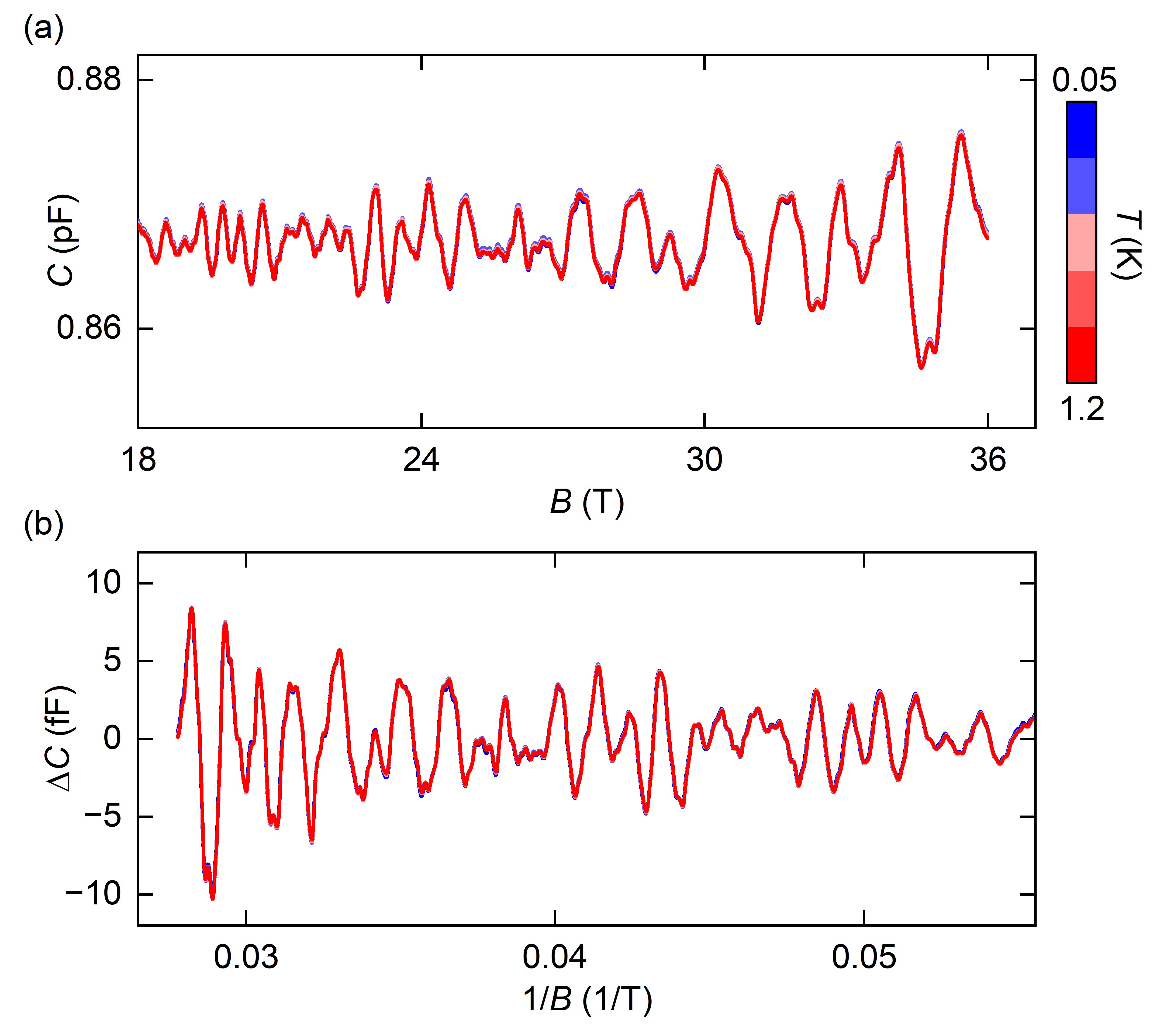}
  \caption{Typical torque magnetometry data recorded at LNCMI Grenoble. (a)~Capacitance, $C$, of the torque magnetometer as a function of $B$ at $\theta = 84^\circ$. (b)~Oscillatory part of the signal as a function of $1/B$. No significant change in oscillation amplitudes is detected in the temperature range between 50\,mK and 1.2\,K.
  }
  \label{fig:High_field_dHvA_overview}
\end{figure}

We additionally performed high-field torque magnetometry experiments in magnetic fields up to 36\,T at the LNCMI Grenoble. The raw capacitance data is shown in Fig.~\ref{fig:High_field_dHvA_overview}(a). The signal exhibits strong oscillations on top of an almost constant background. The oscillatory component of the capacitance is shown in Fig.~\ref{fig:High_field_dHvA_overview}(b). In the experimentally accessible temperature range between 50\,mK and 1.2\,K no significant change in oscillation amplitude was observed. Differences in the magnitude of the capacitance in the overlapping field range between data recorded at TUM and LNCMI are consistent with the different sensitivities of the cantilevers and the different sizes of the samples used.

\subsection{Frequency analysis}

\begin{figure*}
  \includegraphics[width=\textwidth]{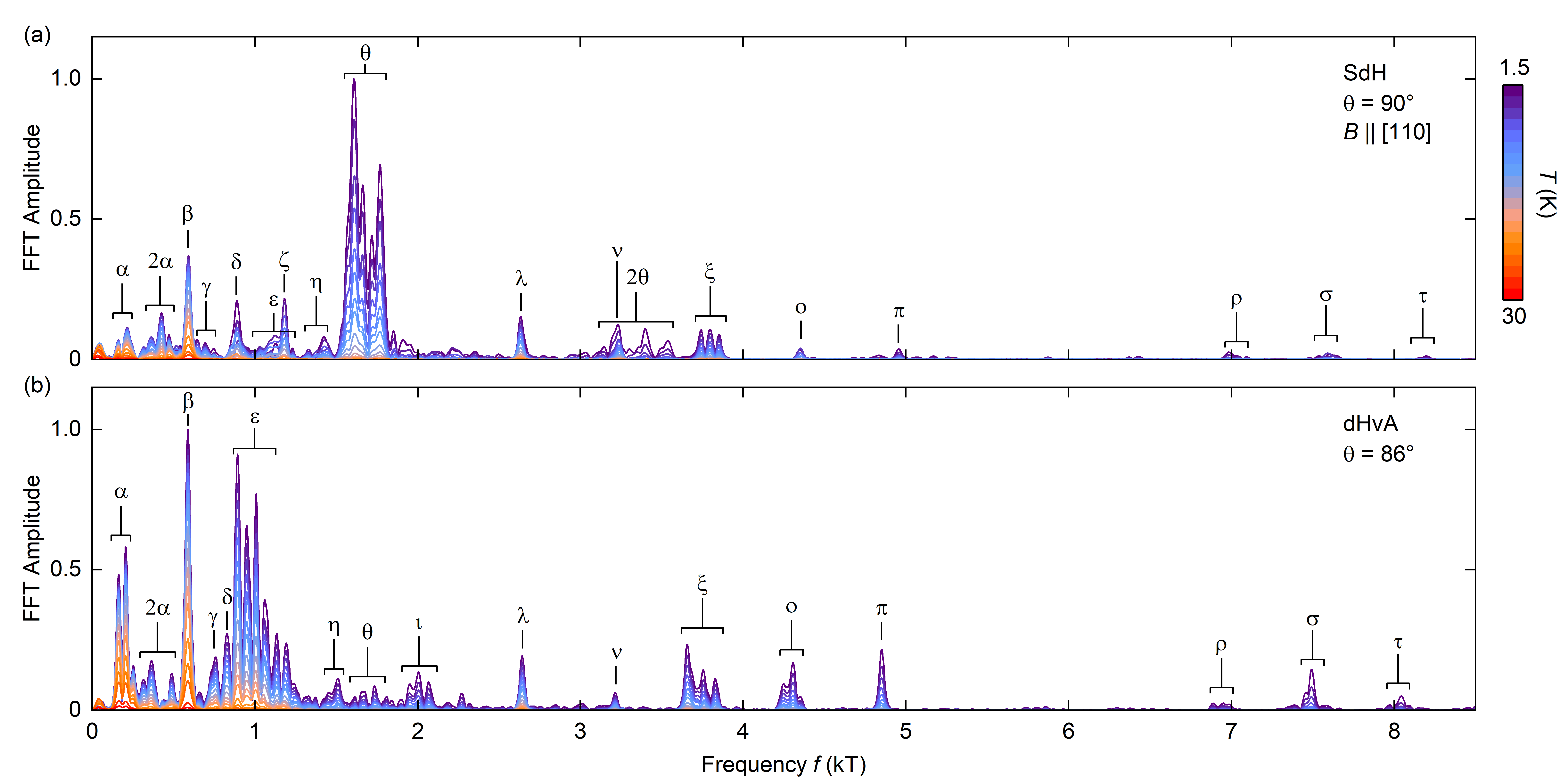}
  \caption{FFT spectra of the quantum oscillations in PdGa. (a)~SdH and (b)~dHvA oscillations recorded at different temperatures between 1.5 and 30\,K close to $\theta=90^\circ$. The dHvA spectra are recorded with $B$ tilted $4^\circ$ from the $[110]$ direction in order to obtain a finite torque. Frequencies are labeled with Greek letters. Neighboring peaks exhibiting similar characteristics are grouped together.}
  \label{fig:FFT_overview}
\end{figure*}

In the following, we analyze the frequency spectrum of the oscillations detected. Fig.~\ref{fig:FFT_overview} shows the FFT spectra of our SdH and torque magnetometry data, analyzed in the field range between 4\,T and 18\,T. Multiple frequencies are detected in both SdH and dHvA oscillations. While most frequency components are observed in both spectra, their relative amplitudes differ strongly. This difference is partially intrinsic and rooted in individual parts of the Fermi surface contributing differently to the magnetization and electrical transport properties. Additionally, the measurement of the dHvA effect through torque magnetometry, which probes the magnetization perpendicular to the applied magnetic field (see Sec.~\ref{sec:ExperimentalMethods}), explains the difference in observed amplitudes.
Both spectra exhibit the same frequency contributions, as denoted by Greek letters. A small mismatch in frequency values may be explained by the slightly different field orientations under which the data were recorded.

The spectra can be grouped into three regimes. The low-frequency regime up to 2.5\,kT comprises many close-lying frequency components, labeled $\alpha$ through $\iota$. Groups of frequencies with similar characteristics are denoted by a single letter with the individual peaks in a group referred to by subscripts in order of increasing frequency (e.g. $\epsilon_1$, $\epsilon_2$, $\dots$). The peaks denoted by $2\alpha$ correspond to the second harmonics of frequencies $\alpha$.
The frequencies labeled $\theta$ dominate the SdH spectrum, indicating that the part of the FS which they originate from contributes strongly to electrical transport. In contrast, the most prominent contributions to the dHvA spectra are the frequencies $\beta$ and $\epsilon$ which we attribute to the strong angular dispersion of the underlying orbits giving rise to a large torque.

The second group of frequencies, denoted $\lambda$ through $\pi$ is located between 2.5\,kT and 5\,kT. Peaks $\lambda$, $\xi_1$-$\xi_3$, and $\pi$ can be clearly identified in both spectra. While only a single peak $o$ is visible in the SdH spectrum, multiple close-by frequencies are detected in the torque signal. Vice versa, multiple low-temperature contributions $2\theta$ and a frequency $\nu$ that varies only weakly as a function of temperature are observed in the SdH oscillations whereas only $\nu$ is identified in the torque spectrum. In contrast to Ref.~\cite{2025_Miyake_JPhysSocJpn}, we observe only a single frequency $\pi$.

In the third regime between 6.5\,kT and 8.5\,kT, three sets of frequencies $\rho$, $\sigma$, and $\tau$ can be discerned in both SdH and dHvA spectra.

\subsection{Angular dispersion}

\begin{figure}
  \includegraphics[width=\columnwidth]{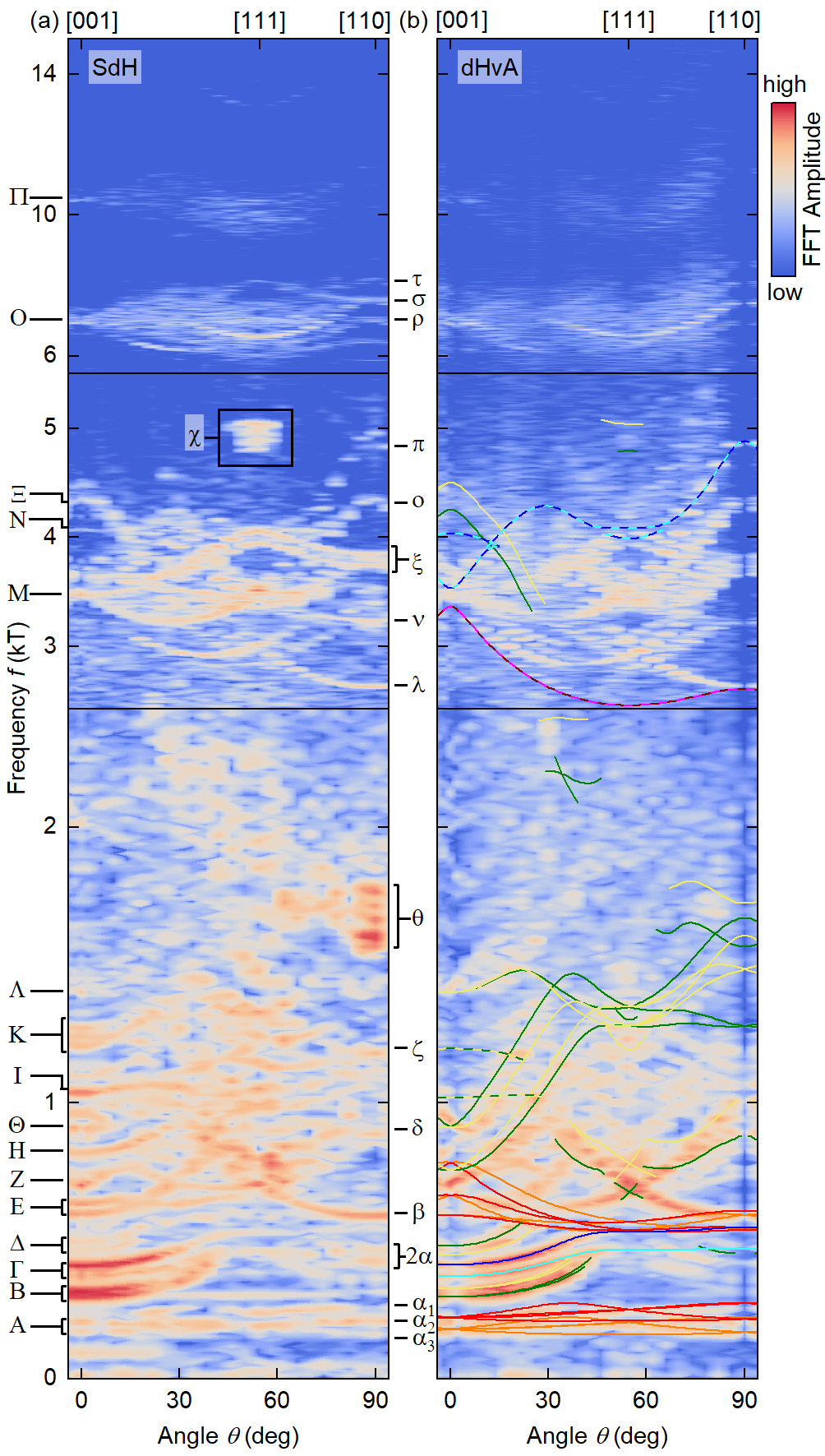}
  \caption{Angular dependence of the quantum oscillation spectra. FFT amplitude (colormap) of the (a)~SdH and (b)~dHvA oscillations recorded at 1.5\,K for different orientations of the applied magnetic field. The observed spectra can be divided into three frequency regimes. Peaks observed at $\theta = 0^\circ$ and $90^\circ$ are labeled with uppercase and lowercase Greek letters, respectively. The lines shown in panel (b) represent predicted frequency branches with colors corresponding to the underlying bands as introduced in Fig.~\ref{fig:bands-FS}.
  }
  \label{fig:FFT_Ang_Dep_Comparison}
\end{figure}

The angular dependence of the quantum oscillations detected was investigated as a function of the orientation of the applied magnetic field in a $(1\bar{1}0)$ rotation plane from $B \parallel [001]$ to $[110]$. The resulting FFT spectra are shown as colormaps in Fig.~\ref{fig:FFT_Ang_Dep_Comparison}. The labeling of frequency contributions at $\theta = 90^\circ$ corresponds to the notation introduced in Fig.~\ref{fig:FFT_overview}. As the complex evolution of frequency branches does not allow for a one-to-one mapping of all branches over the whole angular range, capital Greek letters denote the frequencies detected at $\theta = 0^\circ$.

Overall, the same frequency branches are detected in the SdH and dHvA spectra, albeit with different relative amplitudes. This difference is striking for frequency branches with a low angular dispersion, such as the frequencies $\theta$ or $\chi$, which are strongly suppressed in the torque magnetometry data.
To allow a discussion of the relation between the experimentally detected frequencies and the FS of PdGa, the frequencies calculated from first principles are shown as colored lines in Fig.~\ref{fig:FFT_Ang_Dep_Comparison}(b). Here, the color of individual branches corresponds to the color of the respective FS pockets as depicted in Fig.~\ref{fig:bands-FS}. In the following, we discuss all frequency branches and identify those parts of the spectra that cannot be explained by fundamental frequencies only.

Starting with the lowest branches between 0.15\,kT and 0.25\,kT, a comparison of the frequencies with the DFT predictions arising from the droplet shaped pockets located on the $\Gamma$-R-line yields a good match. They comprise an inner and outer pocket, split by SOC, with eight symmetry related copies in the first BZ. For $B \parallel [001]$, the extremal trajectories on all inner (outer) pockets enclose the same cross-sectional area, explaining the two frequencies $A_1$ and $A_2$ observed. At intermediate angles the different orientations of the pockets lead to a splitting of frequency branches featuring multiple peaks. For $B \parallel [110]$, two distinct frequencies arise from each of the inner and outer pockets. However, the higher frequency arising from the inner pockets is almost identical to the lower frequency from the outer pockets, consistent with the three distinct peaks $\alpha_1$-$\alpha_3$ observed in experiment.

Frequency branches $B_1$ and $B_2$ match very well with extremal trajectories encompassing the pillars of the cage structure centered around the $\Gamma$-point. Because this structure also consists of two nested pockets (an outer yellow and an inner green), two distinct frequency branches with similar dispersion are expected. When increasing $\theta$, the branches rise in frequency and disappear at $\theta \approx 40\degree$ above which orbits around a single pillar are no longer possible.
Similar behavior is observed for the branches labeled $\Delta$ which may be attributed to orbits around the openings in the cage structure. These branches are expected to vanish at $\theta \approx 30\degree$ consistent with experiment.

Two further branches, $\Gamma_1$ and $\Gamma_2$, match the frequencies predicted for extremal trajectories on the cuboid-shaped FS pockets centered around the $\Gamma$-point. The frequency separation indicates a SOC-induced band splitting of $\approx$~60\,meV. Theoretically, these orbits exist for all angles but, due to the curvature of the FS, the largest QO amplitudes are expected for $B \parallel [001]$ consistent with the amplitudes detected experimentally. Additionally, the presence of the frequencies denoted $2\alpha$ makes it difficult to follow the branches originating in $\Gamma_1$ and $\Gamma_2$ close to $B \parallel [110]$.

The experimental frequencies $E$ to $\Lambda$ that evolve into frequencies $\beta$ to $\zeta$ arise from four FS pockets centered around the M-points. Because the pockets around M1, M2, and M3 exhibit different orientations with respect to an applied magnetic field, they give rise to different frequency branches. The lower branches, for which the DFT predictions are shown in red and orange, arise from the inner pockets. In comparison, the upper branches, shown in green and yellow, correspond to extremal orbits on the outer pockets.

We note that the nodal plane degeneracies at the BZ boundary lead to two pairs of intersecting pockets instead of four nested pockets. The latter would give rise to an additional splitting of the frequency branches \cite{2022_Huber_PhysRevLett}.
While the experimentally observed branches in this region fall within the frequency range bounded by the DFT predictions, a significant portion of the spectral weight is observed at frequencies in between and is not fully captured by conventional FS trajectories. This observation may be explained by MB between the inner and the outer FS pockets centered around the M-point.

Further branches predicted at larger frequencies, as depicted in green and yellow, arise from the cage-like FS pockets centered around the $\Gamma$-point. They exist over a limited angular range only.

Close to $\theta = 90^\circ$ we observe a strong contribution from a group of frequencies labeled $\theta$ in the SdH spectra. Their absence in the dHvA spectra is likely related to the small angular dispersion of the underlying frequency branches and the resulting small contribution to the torque signal. We assign this group of frequencies to two frequency branches arising from orbits around the pillars of the cage-like structure around the $\Gamma$-point. We attribute the origin of the large number of frequencies $\theta$ to magnetic breakdown between the two FS pockets. Additionally, frequencies $\theta$ may contain a contribution from the two largest branches arising from the outer M-point pockets.

In the frequency range between 1.8\,kT and 2.5\,kT the FFT spectra exhibit amplitudes above the noise level, cf. Fig.~\ref{fig:FFT_overview}. Except for two frequency branches predicted in a small angular range around $\theta = 35^\circ$ arising from the FS around the $\Gamma$-point, our DFT calculations predict no extremal trajectories and in turn no frequencies in this range. We attribute the spectral weight observed experimentally to higher harmonics of frequencies related to the FS pockets centered around the M-point.

The next higher set of frequency branches comprises a group of frequencies $M$ at $\theta = 0^\circ$ which split into multiple branches with characteristic angular dispersion at intermediate angles evolving into frequencies $\lambda$ to $\pi$ at $\theta = 90^\circ$. A comparison with the predicted frequencies reveals that the frequencies observed experimentally reside between two branches expected from extremal trajectories on the FS centered around the R-point. Similar to the FS around the M-point, it comprises an inner and an outer set of pockets with pockets within a pair mutually intersecting each other at the nodal planes on the BZ boundary. In contrast to the M-point, the R-point exhibits only a single orientation with respect to an applied magnetic field. This leads to two distinct frequency branches only.

We note that Ref.~\cite{2025_Miyake_JPhysSocJpn} does not take the nodal plane degeneracies into account. The authors consequently predict a splitting of each frequency branch in contrast to what we observe in experiment. Again, most of the experimentally detected spectral weight resides between the predicted frequency branches and may be explained by magnetic breakdown trajectories between the inner and the outer set of pockets.

The frequency close to $\theta = 0^\circ$ labeled $N$ originates in an orbit around the outer set of FS sheets around the R-point. Close to $B \parallel [001]$ a set of frequencies $\Xi$ is observed in the SdH spectra.
Taking into account their angular dispersion, we assign these frequencies to extremal trajectories located on the outside of the cage-like FS pockets around the $\Gamma$-point.

Close to $B \parallel [111]$, the SdH spectra feature multiple, nearly dispersionless frequency branches $\chi$ at $\approx$~5\,kT. Their absence in our torque spectra may be explained by their low angular dispersion. In our DFT results, we find extremal trajectories that run diagonally across the cage-like structure around the $\Gamma$-point and exist in this limited angular range only, cf. Fig.~\ref{fig:MB-Gamma-Cage}(a). The frequencies corresponding to orbits on the inner and outer FS pocket match very well with the lowest and highest frequency $\chi$. The detection of further frequencies between them may again be attributed to magnetic breakdown between the two pockets.

Turning to the frequency range above 6\,kT, we find frequency branches that correspond to integer multiples of the branches between $M$ and $\lambda$ to $\pi$. We attribute these branches to magnetic breakdown trajectories between the FS pockets around the R-point involving multiple revolutions. The frequency range of the higher branches narrows for larger frequencies and specific branches exhibit significantly enhanced amplitudes for those parts of the spectrum corresponding to an even number of revolutions (see for example the branch connecting $O$ and $\sigma$). These two effects may be explained by a higher degeneracy of intermediate branches for trajectories involving multiple revolutions and the fact that some trajectories are not closed after a single revolution. We discuss both of these effects in detail in Sec.~\ref{sec:Discussion}.

\subsection{Temperature dependence}

\begin{figure}
  \includegraphics[width=\columnwidth]{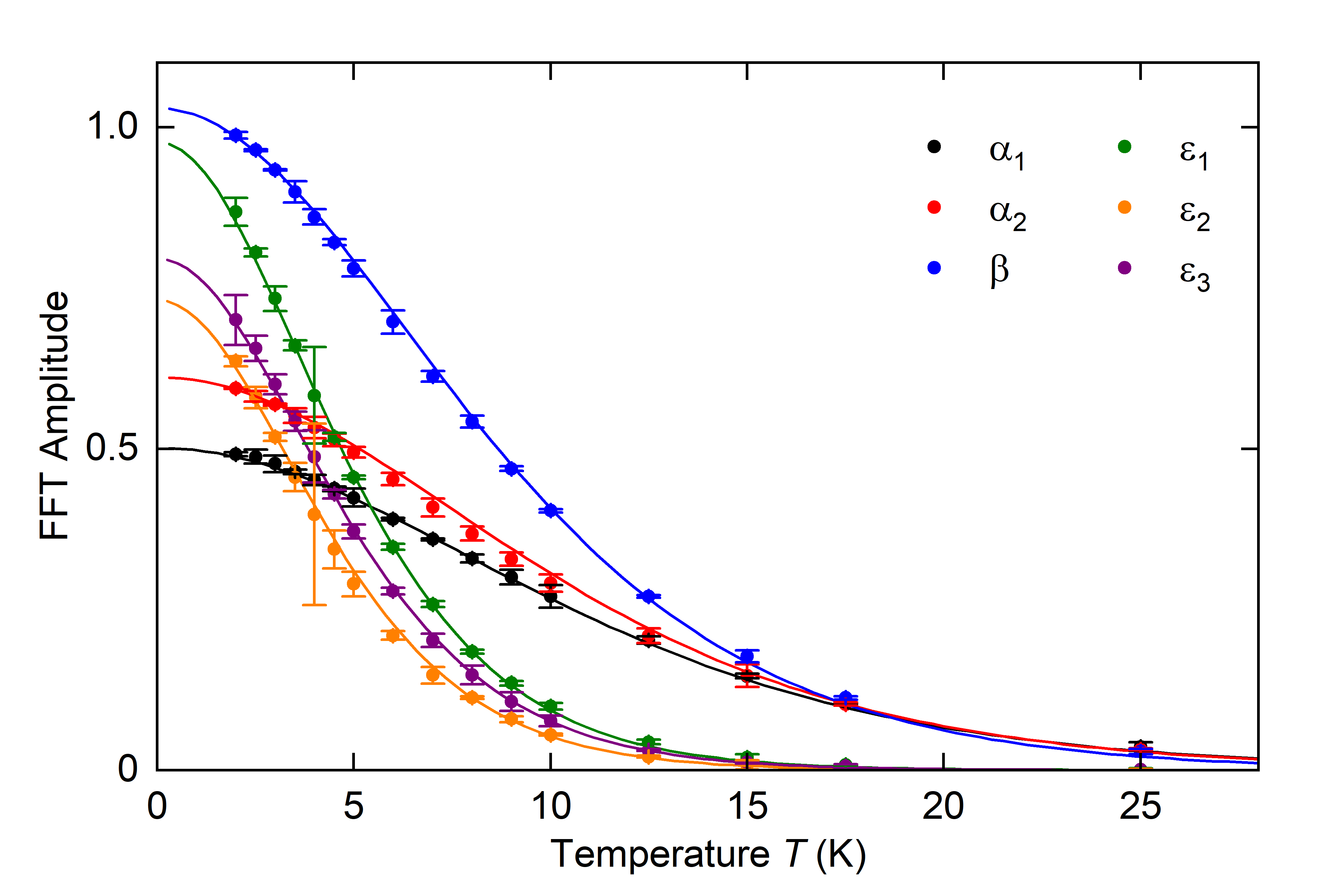}
  \caption{FFT amplitudes of selected dHvA oscillation frequencies as a function of temperature. Solid lines indicate fits with the LK temperature reduction factor $R_T$. The extracted cyclotron masses of all frequencies are given in Tab.~\ref{tab:QO_Analysis}.
  }
  \label{fig:dHvA_T_dependence}
\end{figure}

\begin{figure*}
  \includegraphics[width=\textwidth]{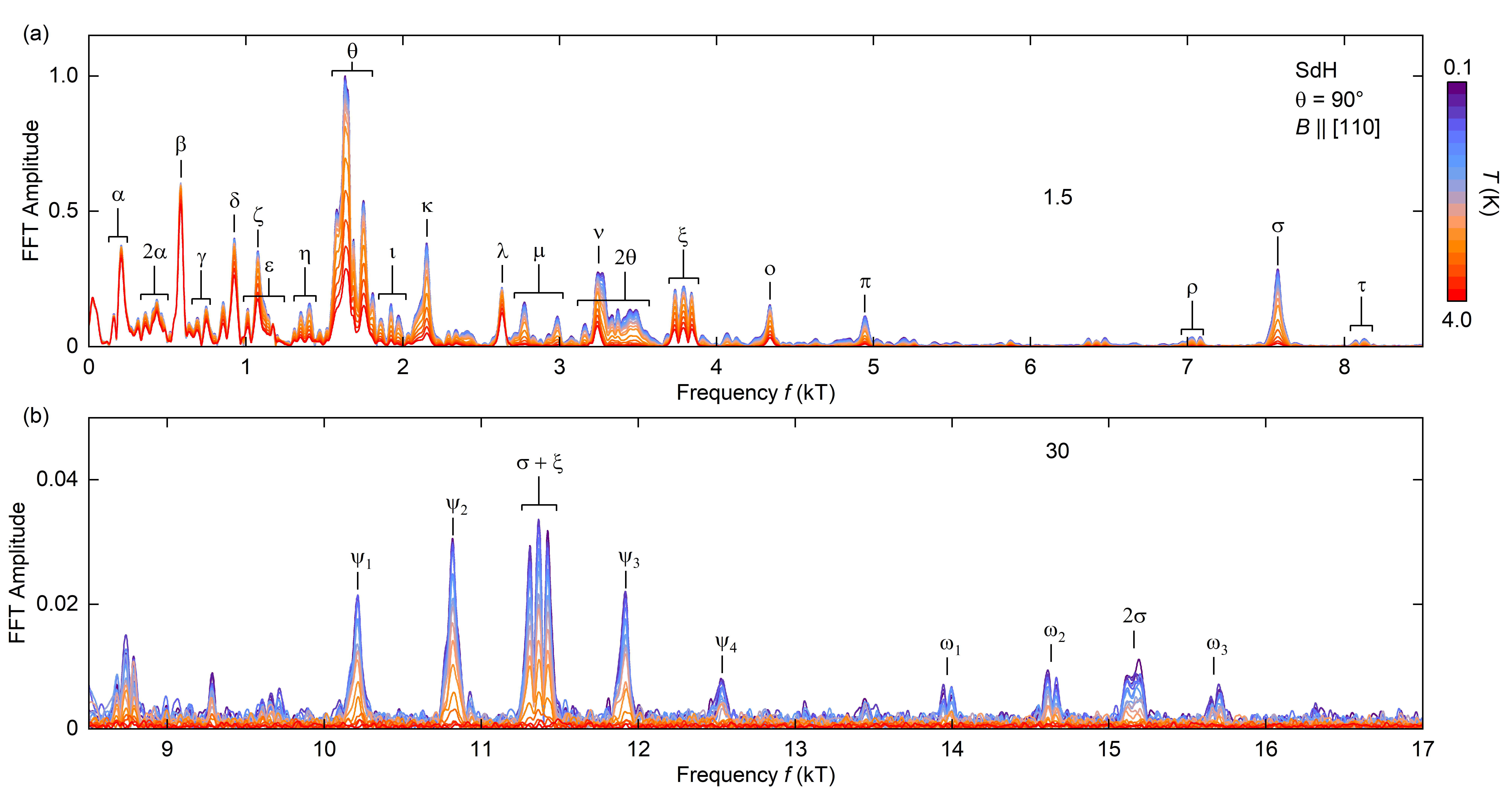}
  \caption{Low-temperature SdH spectra recorded at $\theta=90^\circ$. (a)~Low-frequency regime. Detected frequencies are consistent with the spectra shown in Fig.~\ref{fig:FFT_overview}. Additional features are discussed in the main text. (b)~High-frequency regime containing higher order MB frequencies. }
  \label{fig:SdH_low_T_FFTs}
\end{figure*}

In order to analyze the temperature dependence of the quantum oscillations, the amplitudes of all peaks observed in the FFT spectra were determined by fitting Gaussians to the spectra. Typical data for selected frequencies of the dHvA spectra are shown in Fig.~\ref{fig:dHvA_T_dependence}.
All frequencies follow the LK temperature reduction factor $R_T$. Fits with Eq.\,\eqref{eq:LK-factor} are indicated by solid lines. The extracted cyclotron masses range between 0.1\,$m_e$ and 1.6\,$m_e$, as summarized in Table~\ref{tab:QO_Analysis}.

In order to determine the effective masses of heavier quasiparticle orbits, we additionally recorded SdH spectra with $B\parallel [110]$ at temperatures ranging from 0.1\,K to 4\,K. The corresponding FFT spectra are shown in Fig.~\ref{fig:SdH_low_T_FFTs}. Compared to the high-temperature data shown in Fig.~\ref{fig:FFT_overview}(a), additional maxima emerge. At $\approx$~2\,kT, a group of frequencies labeled $\iota$, which corresponds to a group of frequencies also detected in our torque magnetometry data in Fig.~\ref{fig:FFT_overview}(b), and a strong frequency component $\kappa$ are detected.

While the frequencies at $\iota$ are likely related to second harmonics of fundamental frequencies, the strong amplitude of $\kappa$ indicates a different origin. We attribute this frequency to magnetic breakdown between the cage-like structure around the $\Gamma$-point and two of the droplets nested inside it. At $\approx$~3\,kT, a group of frequencies $\mu$ emerges and the frequency $\nu$ is superposed by multiple close-by frequencies $2\theta$.
At 7.6\,kT, the frequency component $\sigma$ grows significantly in amplitude below 4\,K and dominates the FFT spectrum in the high-frequency range at 0.1\,K.

At frequencies above 10\,kT, shown in Fig.~\ref{fig:SdH_low_T_FFTs}(b), several additional peaks emerge in the spectra at temperatures below 4\,K. The group of frequencies labeled $\sigma + \xi$ is similar in structure and almost exactly at three times the frequency of $\xi$. The separation of the peaks, however, remains constant for $\xi$ and $\sigma + \xi$ indicating that they are not simply harmonics of $\xi$. Adjacent to $\sigma + \xi$ there are evenly spaced frequency contributions $\psi_i$ with a separation of $\approx 0.6$\,kT. At larger frequencies just above 15\,kT, a peak corresponding to $2\sigma$ is detected which again is surrounded by evenly spaced frequency contributions $\omega$ separated by $\approx 0.6$\,kT. This part of the spectrum may be attributed to magnetic breakdown involving multiple revolutions with distinct trajectories on the FS pockets centered around the R-point.
The effective masses inferred from all of our temperature dependent spectra are summarized in Table~\ref{tab:QO_Analysis}.

\begin{table*}
\begin{ruledtabular}
\begin{tabular}{cccccccccc}
Peak &  \multicolumn{2}{c}{SdH high $T$} & \multicolumn{2}{c}{dHvA high $T$} & \multicolumn{2}{c}{SdH low $T$} & \multicolumn{2}{c}{DFT} & Origin\\
$B \parallel [110]$ & $f$ (kT) & $m^*$ ($m_{e}$) & $f$ (kT) & $m^*$ ($m_{e}$) & $f$ (kT) & $m^*$ ($m_{e}$) & $f$ (kT) & $m^*$ ($m_{e}$) & FS pocket\\
\hline
$\alpha_1$ &   &   & 0.162(5) & 0.170(10) &   &   & 0.17  & 0.15  & $\Gamma$-R droplet\\
$\alpha_2$ & 0.215(6) & 0.190(10) & 0.207(5) & 0.180(11) & 0.207(6) & 0.190(11) & 0.22  & 0.16 & $\Gamma$-R droplet \\
$2\alpha_1$ &   &   & 0.363(7) & 0.330(20) &   &   & 0.34  & 0.30 & $\Gamma$-R droplet (harm.)\\
$2\alpha_2$ & 0.426(10) & 0.330(20) &   &   & 0.434(13) & 0.290(20) & 0.44  & 0.32  & $\Gamma$-R droplet (harm.)\\
$2\alpha_3$ & 0.471(7) & 0.200(21) &   &   &   &   &   &  & $\Gamma$-R droplet (harm.) \\
$\beta$ & 0.592(10) & 0.230(13) & 0.587(5) & 0.210(12) & 0.585(11) & 0.190(11) & 0.59  & 0.22  & $\Gamma$-cube\\
$\gamma$ &   &   & 0.753(6) & 0.340(21) & 0.749(11) & 0.350(23) &   &  & M-point (MB) \\
$\delta$ & 0.891(14) & 0.45(5) & 0.828(6) & 0.390(22) & 0.857(12) & 0.420(26) &   &  & M-point (MB) \\
$\epsilon_1$ &   &   & 0.893(5) & 0.370(20) &   &   &   &  & M-point (MB) \\
$\epsilon_2$ &   &   & 0.953(7) & 0.410(24) & 0.928(12) & 0.330(21) &   &  & M-point (MB) \\
$\epsilon_3$ &   &   & 1.007(10) & 0.380(21) & 1.012(12) & 0.420(25) &   &  & M-point (MB) \\
$\epsilon_4$ &   &   & 1.064(11) & 0.390(25) & 1.079(16) & 0.450(24) &   &  & M-point (MB) \\
$\epsilon_5$ &   &   & 1.131(8) & 0.520(29) &   &   &   &  & M-point (MB) \\
$\zeta$ & 1.184(15) & 0.53(8) & 1.193(12) & 0.510(30) &   &   &   &  & M-point (MB) \\
$\eta_2$ &   &   &   &   & 1.406(14) & 0.72(4) &   & & M-point (MB)  \\
$\theta_1$ & 1.610(19) & 0.69(4) &   &   & 1.591(24) & 0.59(4) & 1.61  & 0.66 & $\Gamma$-cage \\
$\theta_2$ & 1.651(27) & 0.63(4) &   &   & 1.641(19) & 0.630(33) & 1.67  & 0.69 & $\Gamma$-cage \\
$\theta_3$ &   &   &   &   & 1.752(20) & 0.630(33) &   &  & $\Gamma$-cage \\
$\theta_4$ & 1.770(20) & 0.70(4) &   &   & 1.805(21) & 0.76(5) & 1.72  & 0.67 & $\Gamma$-cage \\
$\theta_5$ & 1.85(4) & 1.00(8) &   &   & 1.927(23) & 0.89(6) &   &  & $\Gamma$-cage \\
$\iota_1$ &   &   & 2.002(9) & 0.84(5) &   &   &   & & M-point (harm.) \\
$\iota_2$ &   &   & 2.067(9) & 0.74(4) &   &   &   & & M-point (harm.) \\
$\kappa$ &   &   &   &   & 2.154(26) & 0.87(5) &   & & $\Gamma$-cage, droplet (MB) \\
$\lambda$ & 2.639(27) & 0.440(26) & 2.642(6) & 0.440(26) & 2.633(26) & 0.390(22) & 2.61  & 0.42  & R-point\\
$\mu_1$ &   &   &   &   & 2.777(23) & 1.11(6) &   &  & harm. \\
$\mu_2$ &   &   &   &   & 2.986(28) & 1.12(6) &   &  & harm. \\
$\nu$ & 3.227(33) & 0.80(6) &  3.212(6) & 0.550(32)  & 3.259(17) & 0.86(7) &  3.19 & 0.53 & R-point (MB) \\
$2\theta_2$ & 3.402(33) & 1.54(11) &   &   &   &   &   &  & $\Gamma$-cage (harm.) \\
$2\theta_3$ & 3.534(29) & 1.39(8) &   &   & 3.47(4) & 1.21(7) &   &  & $\Gamma$-cage (harm.) \\
$\xi_1$ & 3.76(5) & 0.65(4) & 3.662(8) & 0.610(34) & 3.75(4) & 0.630(35) & 3.68  & 0.65 & R-point (MB) \\
$\xi_2$ & 3.80(4) & 0.62(4) &   &   & 3.796(34) & 0.590(35) & 3.73  & 0.65 & R-point (MB) \\
$\xi_3$ &   &   & 3.823(9) & 0.68(4) & 3.826(27) & 0.65(4) &  3.79 &  0.65 & R-point (MB)\\
$o$ & 4.35(4) & 0.62(8) &   &   & 4.34(4) & 0.70(4) & 4.32  & 0.76 & R-point (MB) \\
$\pi$ & 4.96(4) & 0.90(6) & 4.851(10) & 0.86(5) & 4.95(4) & 1.06(8) & 4.88  & 0.87 & R-point \\
$\rho_1$ &   &   & 6.886(21) & 1.08(9) & 7.04(6) & 0.89(9) &  6.87 &  1.18 & R-point (MB 2$^\mathrm{nd}$)\\
$\rho_2$ & 7.00(5) & 1.32(14) & 6.95(6) & 1.14(17) & 7.09(6) & 1.01(7) &  6.92 &  1.18 & R-point (MB 2$^\mathrm{nd}$) \\
$\sigma$ & 7.54(6) & 2.3(4) & 7.450(13) & 1.40(10) & 7.58(6) & 1.12(6) & 7.49  & 1.29 & R-point (MB 2$^\mathrm{nd}$) \\
$\tau_1$ &   &   & 7.969(11) & 1.08(14) & 8.08(5) & 1.16(8) &  8.01 & 1.40 & R-point (MB 2$^\mathrm{nd}$)\\
$\tau_2$ & 8.20(6) & 1.44(17) & 8.046(26) & 1.50(11) & 8.14(6) & 1.18(8) &  8.06 &  1.40 & R-point (MB 2$^\mathrm{nd}$) \\
$\psi_1$ &   &   &   &   & 10.22(7) & 1.41(12) & 10.07  &  1.84 & R-point (MB 3$^\mathrm{rd}$)\\
$\psi_2$ &   &   &   &   & 10.83(7) & 1.49(10) &  10.62  &  1.89 & R-point (MB 3$^\mathrm{rd}$)\\
$\sigma+\xi_1$ &   &   &   &   & 11.31(6) & 1.63(11) &  11.17 &  1.94 & R-point (MB 3$^\mathrm{rd}$)\\
$\sigma+\xi_2$ &   &   &   &   & 11.38(7) & 1.60(11) & 11.22  &  1.94 & R-point (MB 3$^\mathrm{rd}$)\\
$\sigma+\xi_3$ &   &   &   &   & 11.44(7) & 1.53(10) &  11.27 &  1.94 & R-point (MB 3$^\mathrm{rd}$)\\
$\psi_3$ &   &   &   &   & 11.92(7) & 1.65(12) &  11.72 &  1.99 & R-point (MB 3$^\mathrm{rd}$)\\
$\omega_2$ &   &   &   &   & 14.02(8) & 1.4(4) & 13.88 &  1.48 & R-point (MB 4$^\mathrm{th}$)\\
$\omega_3$ &   &   &   &   & 14.62(7) & 1.8(4) & 14.43 &  2.03 & R-point (MB 4$^\mathrm{th}$)\\
$2\sigma$ &   &   &   &   & 15.165(13) & 2.1(2) &  14.98 &  2.58 & R-point (MB 4$^\mathrm{th}$)\\
$\omega_4$ &   &   &   &   & 15.69(15) & 1.52(27) &  15.53 &  3.13 & R-point (MB 4$^\mathrm{th}$)\\
\end{tabular}
\end{ruledtabular}
\caption{Summary of the experimentally observed frequencies $f$ and extracted cyclotron masses $m^*$ for magnetic field along the [110] direction (at $\theta = 86^\circ$ for the dHvA data). A comparison with values predicted by DFT calculations and the Fermi surface pockets on which the trajectories reside are given in the rightmost columns. Higher harmonics of fundamental frequencies are denoted by (harm.). Magnetic breakdown orbits are indicated as (MB) and the breakdown order is given for higher order MB trajectories, cf. Sec.~\ref{sec:Discussion_R-point}.}
\label{tab:QO_Analysis}
\end{table*}

\section{Discussion}\label{sec:Discussion}

As demonstrated above, nearly all fundamental frequency branches predicted by DFT are detected in our experiments. The match between the extremal cross sections obtained from the DFT band structure data and the frequencies measured in experiment is excellent. However, there are also additional frequency branches observed experimentally which cannot be explained by conventional quasiparticle trajectories on a single FS pocket. They may be attributed to magnetic breakdown (MB) trajectories connecting different FS pockets at MB junctions at which the individual pockets exhibit a minimal separation in $k$-space. The complex FS of PdGa, comprising multiple mutually intersecting FS pockets at the M- and R-point, as well as nested FS pockets centered around the $\Gamma$-point, features many regions in $k$-space for which significant MB is expected.
We calculated the MB frequencies and probabilities with the procedure outlined in Sec.~\ref{sec:ExperimentalMethods} and compare our results to the experimentally observed frequency branches in the following.

\subsection{MB between FS pockets at the \texorpdfstring{$\Gamma$}{Gamma}-point}

The FS of PdGa centered around the $\Gamma$-point, shown in Figs.~\ref{fig:bands-FS}(c) and \ref{fig:bands-FS}(d), comprises multiple nested pockets which appear in pairs of similar shape each split by SOC. We analyzed the distance between the individual FS pockets in order to identify potential MB junctions. Note that the separation between the sets of pockets at the $\Gamma$-point is considerably larger than the separation within a pair, thus rendering MB between the sets of pockets unlikely except for a special case at $B \parallel [110]$.

For the centermost cuboid-shaped pockets, depicted in blue and cyan, and the droplet-shaped pockets along the $\Gamma$-R-line, depicted in red and orange, the $k$-space separation is sufficiently large to suppress MB in the field range of our experiments. Correspondingly, the experimentally observed frequency branches $\Gamma_{1,2}$ and $A_{1,2}$ to $\alpha_{1,2,3}$ are explained well by the predicted branches arising from these pockets without MB, cf. Fig.~\ref{fig:FFT_Ang_Dep_Comparison}(b).

\begin{figure}
  \includegraphics[width=\columnwidth]{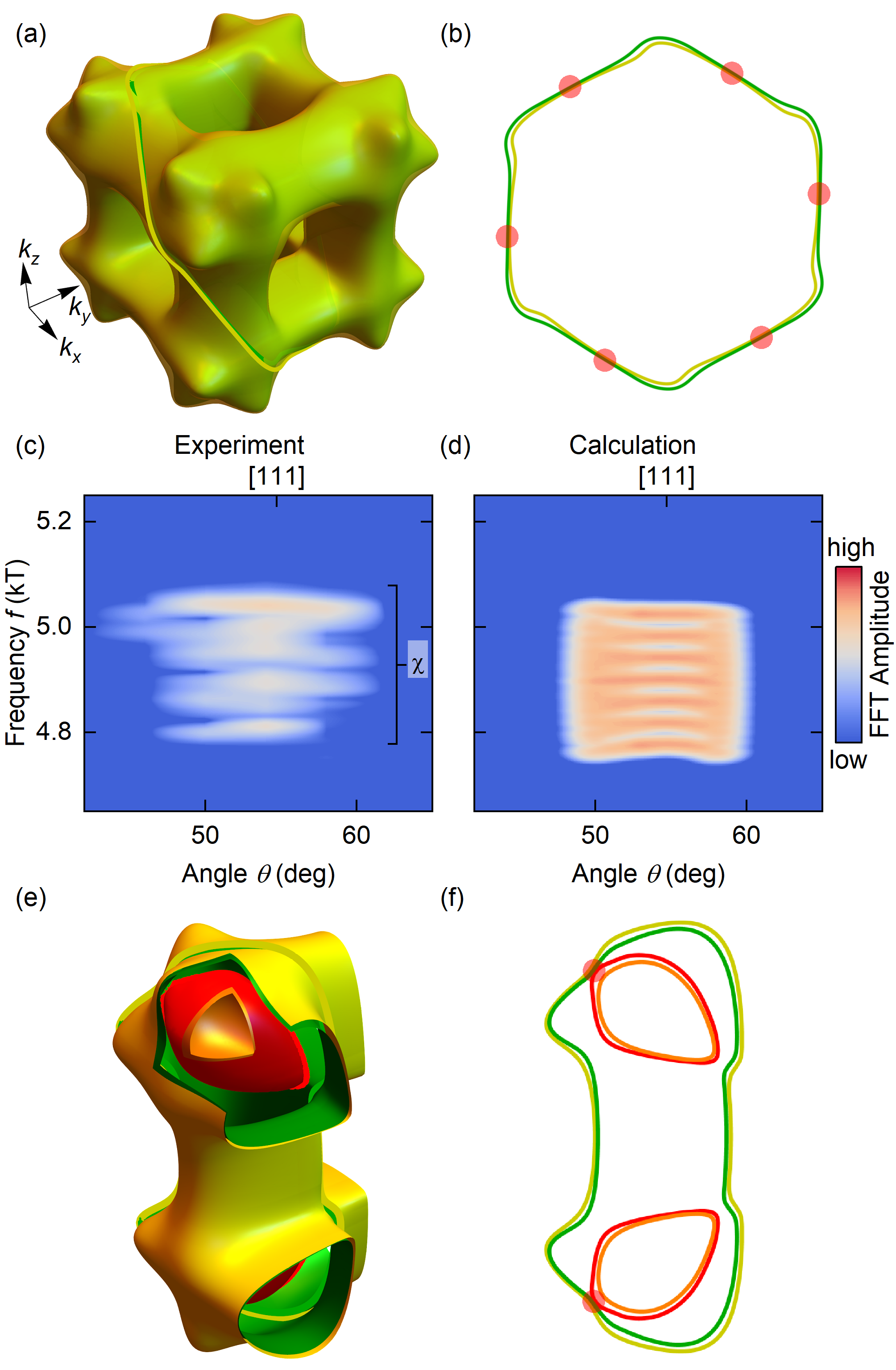}
  \caption{Magnetic breakdown involving the cage-like FS sheets centered around $\Gamma$. (a)~FS pockets centered around the $\Gamma$-point comprising two nested cage-like structures. Extremal orbits for $B \parallel [111]$ are drawn as colored lines. (b)~Extremal trajectories perpendicular to the [111] direction. MB junctions are indicated by colored circles. (c)~Measured frequency branches $\chi$, as shown in Fig.~\ref{fig:FFT_Ang_Dep_Comparison}(a). (d)~Simulated MB frequencies for comparison. (e)~Cutout of one of the pillars of the cage with droplet-shaped FS pockets in its corners. Extremal orbits for $B \parallel [110]$ are drawn as colored lines. (f)~Extremal trajectories around the pillars perpendicular to the [110] direction. The gap of the MB junctions highlighted by colored circles sensitively depends on the field direction with a minimum for fields applied exactly along [110].}
  \label{fig:MB-Gamma-Cage}
\end{figure}

In contrast, the cage-like structure comprising the nested pockets depicted in yellow and green exhibits regions for which the $k$-space separation is very small. This permits magnetic breakdown at orientations of the applied magnetic field at which the extremal trajectories traverse through these regions. This behavior is observed for a magnetic field applied along the [111] direction. The extremal trajectories on the FS pockets depicted in yellow and green perpendicular to the [111] direction are shown in Figs.~\ref{fig:MB-Gamma-Cage}(a) and \ref{fig:MB-Gamma-Cage}(b).

When neglecting magnetic breakdown, they give rise to two frequencies slightly above and below 5\,kT. In experiment, we observe these two frequencies in our SdH spectra and assign them to the group of frequencies $\chi$, cf. Fig.~\ref{fig:MB-Gamma-Cage}(c).
The outermost frequency branches $\chi$ match the predicted frequencies from extremal orbits around the two cage-like pockets very well. However, the experimental spectra exhibit additional frequency branches we attribute to MB orbits connecting the two nested pockets at the breakdown junctions highlighted in Fig.~\ref{fig:MB-Gamma-Cage}(b).

We calculated the MB frequencies and probabilities of the orbits for magnetic fields applied close to the [111] direction. They are shown as a colormap in Fig.~\ref{fig:MB-Gamma-Cage}(d). The experimental spectra match the simulated behavior of the MB branches very well.

Additionally, the cage-like pockets exhibit extremal cross sections around the pillars on their edges perpendicular to the [110] direction. These orbits correspond to frequencies $\theta$ observed in our SdH-spectra, cf. Fig.~\ref{fig:FFT_Ang_Dep_Comparison}(a). Again, the outermost frequencies $\theta$ match the values predicted by DFT very well. We attribute the additional frequency contributions to magnetic breakdown orbits between the two FS sheets comprising the cage structure.

A strong frequency contribution $\kappa$ is observed in our low-temperature SdH spectra, shown in Fig.~\ref{fig:SdH_low_T_FFTs}. We attribute this frequency to MB orbits connecting trajectories on the cage-like structure with the droplets located inside it. A cutout of the FS and the corresponding extremal cross sections are depicted in Figs.~\ref{fig:MB-Gamma-Cage}(e) and \ref{fig:MB-Gamma-Cage}(f). The frequency and effective mass of $\kappa$, given in Table~\ref{tab:QO_Analysis}, closely matches the combination $\theta_3+2\alpha_2$ indicating a breakdown orbit that encloses the pillar and both droplets as its origin.
The size of the gap at the breakdown junctions, depicted in Fig.~\ref{fig:MB-Gamma-Cage}(f), sensitively depends on the orientation of the magnetic field with a minimum for extremal cross sections exactly perpendicular to the [110] direction.

The presence of $\kappa$ in our low-temperature SdH spectra, shown in Fig.~\ref{fig:SdH_low_T_FFTs}, and its absence in the high-temperature SdH spectra, shown in Fig.~\ref{fig:FFT_overview}(a), may therefore be explained by a small tilt of the magnetic field away from the [110] direction for the latter.
The small separation between orbits on the outer droplet-shaped sheet and the inner cage-like structure and its sensitivity to the angle of the applied magnetic field may be explained by a Weyl-point between bands 2 and 3 located along the $\Gamma$-R line that resides $\approx 30$\,meV above the Fermi energy, cf. Fig.~\ref{fig:bands-FS}(a).

\subsection{MB between FS pockets at the M-point}

\begin{figure}
  \includegraphics[width=\columnwidth]{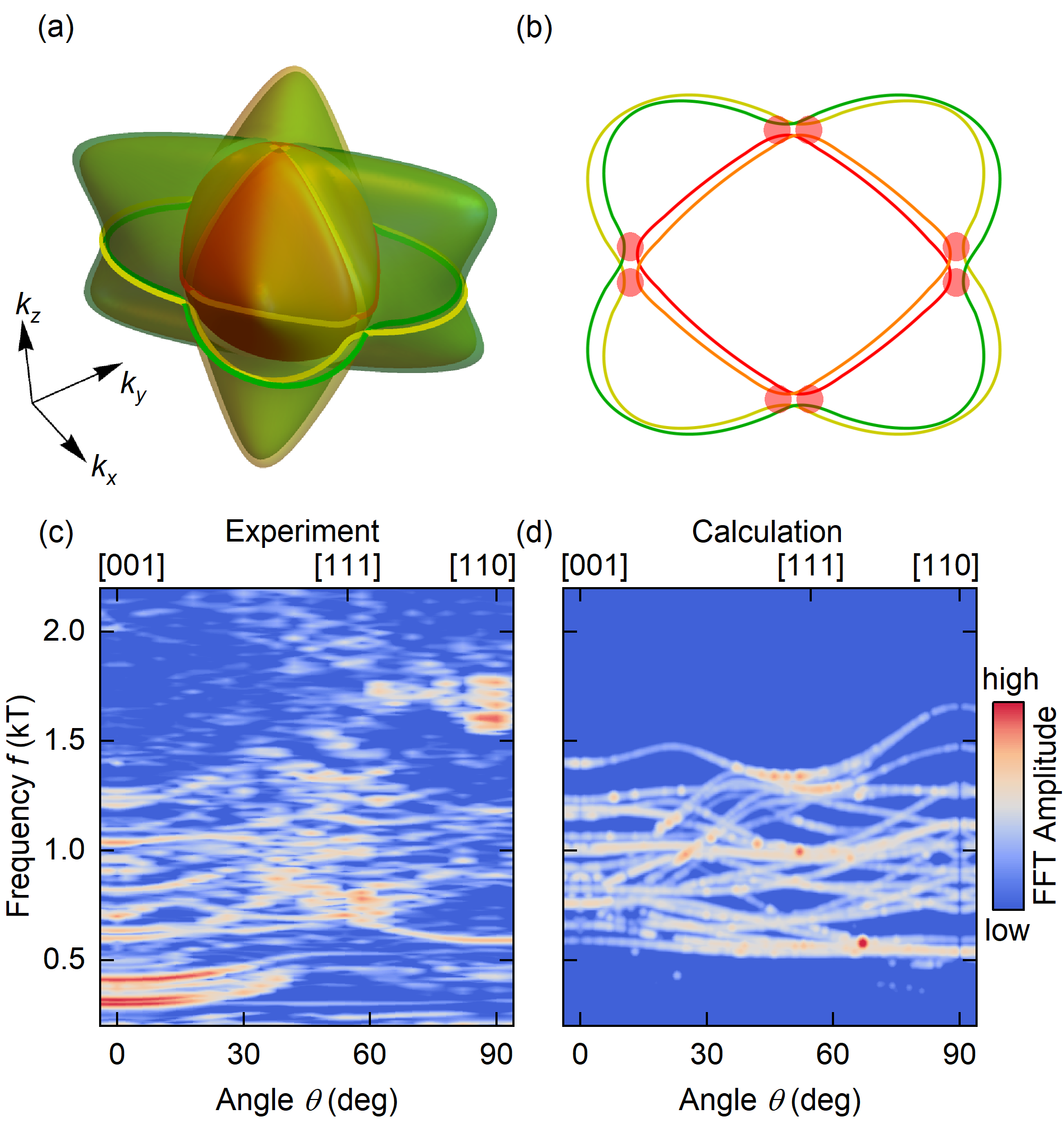}
  \caption{Magnetic breakdown at the M-point. (a)~FS centered around the M-point comprising two sets of intersecting pockets. Extremal orbits for $B \parallel [001]$ are drawn as colored lines. (b)~Extremal trajectories perpendicular to the [001] direction with MB junctions highlighted by colored circles. (c),~(d)~Comparison between the measured and predicted frequency spectra. Note that the experimental spectra contain frequency contributions arising from other parts of the FS while the calculation only includes frequencies related to the FS pockets located around the M-point.}
  \label{fig:MB-M-point}
\end{figure}

The FS centered around the M-point consists of an inner and an outer pair of pockets for which pockets within a pair intersect each other at the nodal planes on the BZ boundary, cf. Figs.~\ref{fig:bands-FS}(c) and \ref{fig:bands-FS}(d). The three orientations of these pockets with respect to an applied magnetic field are expected to give rise to a rich frequency spectrum in the range between 0.5\,kT and 1.5\,kT, see Fig.~\ref{fig:FFT_Ang_Dep_Comparison}(b). Again, we observe additional spectral weight at frequencies between the predicted frequency branches which may be explained by MB between the inner and the outer pairs of pockets. We calculated the breakdown frequencies and the associated amplitudes for all three orientations of the M-point pockets with the procedure outlined in Sec.~\ref{sec:Methods_MB} and show them as a colormap in Fig.~\ref{fig:MB-M-point}(d).

A comparison with the experimentally observed frequencies is in qualitative agreement with additional spectral weight between the frequency branches predicted for the FS pockets without MB. The coexistence of conventional and MB branches in our spectra indicates only partial MB in the experimentally accessible field range.
Some calculated MB branches and their expected amplitudes match the experimental observations very well. However, the complex spectrum in this frequency range, which additionally includes frequency contributions from other parts of the FS, prohibits a one-to-one assignment of the experimentally detected frequencies to the calculated MB orbits.
Nevertheless, the good agreement between the experimental and calculated spectra confirm the calculated geometry of the FS centered around the M-point. This shows that it consist of two pairs of nested pockets separated by small SOC-induced gaps.

\subsection{MB between FS pockets at the R-point}\label{sec:Discussion_R-point}

Similar to the M-point, the FS of PdGa around the R-point comprises two sets of pockets, cf. Figs.~\ref{fig:bands-FS}(c) and \ref{fig:bands-FS}(d), which again are mutually intersecting within a pair due to the nodal plane degeneracies on the BZ boundary. The larger size of the pockets centered around the R-point as compared to the M-point pockets results in larger QO frequencies in the range between 2.5\,kT and 5\,kT.

\begin{figure}
  \includegraphics[width=\columnwidth]{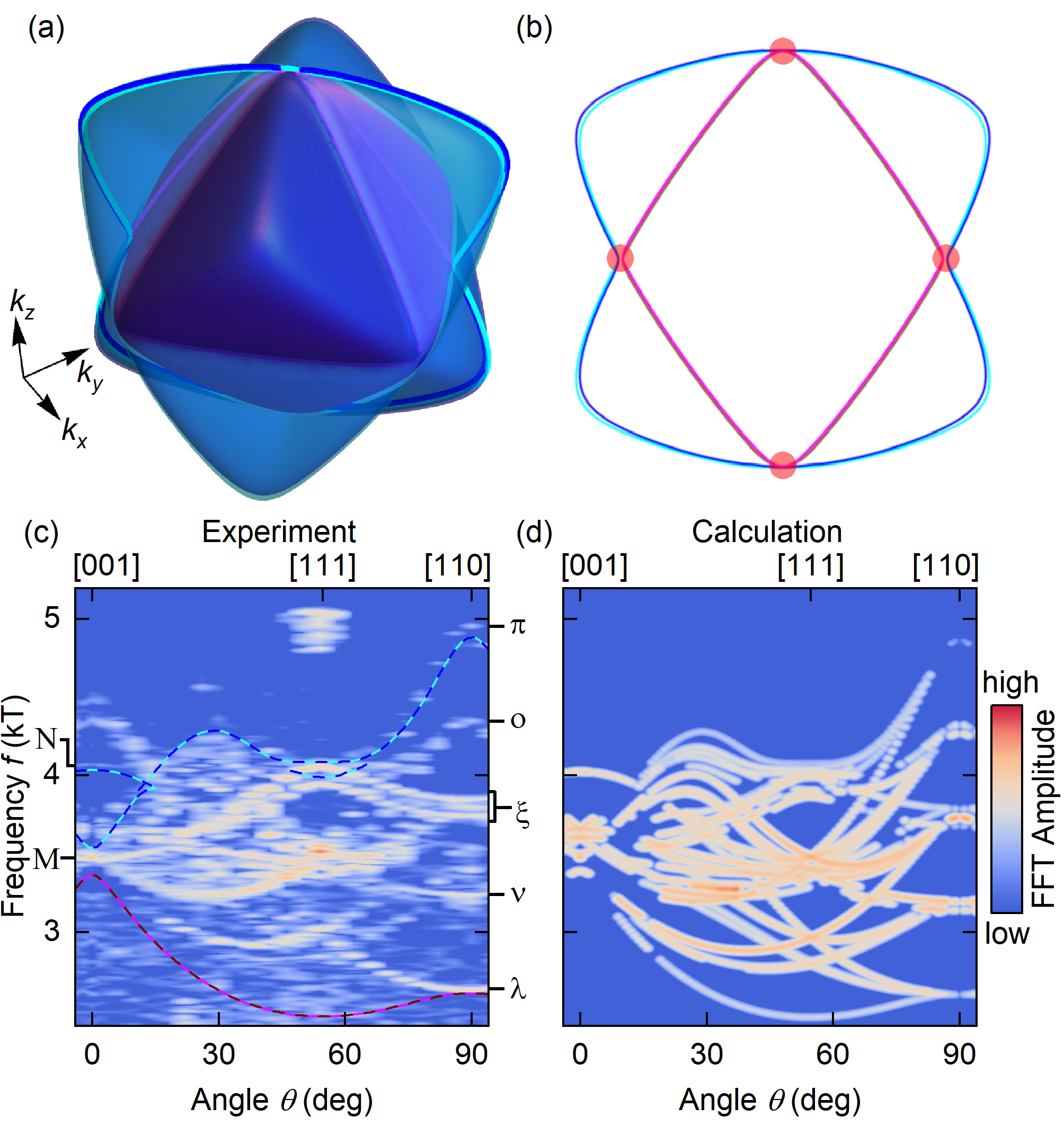}
  \caption{Magnetic breakdown at the R-point. (a)~FS centered around the R-point comprising two sets of intersecting pockets. Extremal orbits for $B \parallel [110]$ are drawn as colored lines. (b)~Extremal trajectories for fields along the [110] direction. MB junctions are highlighted by colored circles. (c)~Experimental SdH spectra in the frequency region containing the R-point frequencies. Dashed lines represent the predicted QO frequencies without MB. (d)~Calculated MB branches for comparison.}
  \label{fig:MB-R-point}
\end{figure}

The existence of only a single R-point, as opposed to the three different M-points, simplifies the QO spectrum significantly. The fundamental frequency branches predicted comprise a lower branch arising from the inner pair of pockets and a higher branch arising from the outer pair of pockets, cf. Fig.~\ref{fig:FFT_Ang_Dep_Comparison}(b). Due to multiple extremal cross sections on the outer FS pockets perpendicular to the [001] and [111] directions, the higher frequency branch splits into two branches in the vicinity of these high-symmetry orientations.
Upon comparison with the experimental observations, we find that both frequency branches predicted are only weakly visible in our spectra. Instead, most of the spectral weight in this frequency range is found between these two frequencies. In contrast to the frequency regime containing the frequencies related to the M-point, no other contributions are expected (apart from two branches arising from the FS around the $\Gamma$-point close to the [001] direction).
The frequency branches observed originate in MB between the inner and the outer pair of pockets centered around the R-point which we analyze in detail in the following.

Figure~\ref{fig:MB-R-point} shows a comparison between the experimentally observed frequency branches and the calculated frequencies arising from the R-point pockets including MB. Consistent with experiment, most of the spectral weight is found between the two frequency branches expected from trajectories without MB. The outer branches are still visible in the predicted and the experimentally observed frequency spectra, implying only partial MB at the breakdown junctions. This situation contrasts the complete magnetic breakdown observed between the R-point pockets in CoSi \cite{2022_Huber_PhysRevLett, 2022_Guo_NatPhys} indicating that quasi-symmetries do not stabilize full MB at the FS around the R-point in PdGa.

Further features of our experimental data, shown in Fig.~\ref{fig:MB-R-point}(c), are reproduced well by the theoretically predicted spectra, as depicted in Fig.~\ref{fig:MB-R-point}(d). The frequency branches merge into only a few frequencies when the magnetic field is applied along the [001] direction. Two of them, $M$ and $N$, are identified clearly in our experimental data.
Close to the [111] direction, multiple branches intersect at $\approx$~3.5\,kT, giving rise to a broad regime with large weight in both our experimental and the predicted spectra.
For the [110] direction, five distinct sets of frequencies are predicted, corresponding to the frequency branches $\lambda$ to $\pi$ observed experimentally. They are evenly spaced with a separation of $\approx$~0.6\,kT corresponding to the area enclosed by one `lobe' on the outer trajectories, cf. Fig.~\ref{fig:MB-R-point}(b).

We also observe the influence of the SOC-induced splitting of the bands into two pairs that leads to two slightly different areas of the lobes depending on which band is involved in the MB trajectory. If no lobe (or all four lobes) are included, no splitting of the frequencies is expected. If one (or three) lobes are included, two distinct frequencies with a splitting of $\approx$~0.05\,kT are expected from the different trajectories. If two lobes are included, we expect a threefold splitting depending on which combination of bands is included in the MB trajectory.

Experimentally, we observe this splitting most prominently for the set of frequencies $\xi$, which was also noted in Ref.~\cite{2025_Miyake_JPhysSocJpn}, corresponding to trajectories involving two lobes. In the spectra shown in Figs.~\ref{fig:FFT_overview} and \ref{fig:SdH_low_T_FFTs} we clearly resolve three distinct peaks with frequencies separated by $\approx$~0.05\,kT. This is consistent with the calculated band splitting of $\approx$~25\,meV. We note that the SOC-induced splitting of the inner trajectories could in principle also give rise to a splitting of frequencies. However, because the difference of the areas enclosed by the inner trajectories is small as compared to the splitting of the trajectories on the lobes, we cannot resolve this splitting in our experimental spectra.

While the angular dispersion of the frequency branches shows excellent agreement between experiment and theory, discrepancies exist with regard to the predicted amplitudes. For instance, two clear frequency branches connecting the peak at $M$ via a maximum at $\theta=[111]$ with the group of peaks denoted $\xi$ exhibit a larger amplitude in our experiments than predicted. We attribute this difference in amplitudes to the consideration of breakdown between two of the FS pockets only (one inner and one outer one) instead of the actual four. This simplification has negligible effect on the predicted frequencies because the FS pockets within a pair are related by symmetry but may influence the relative amplitudes since they sensitively depend on the size of the gap at the breakdown junctions.

\begin{figure*}
  \includegraphics[width=\textwidth]{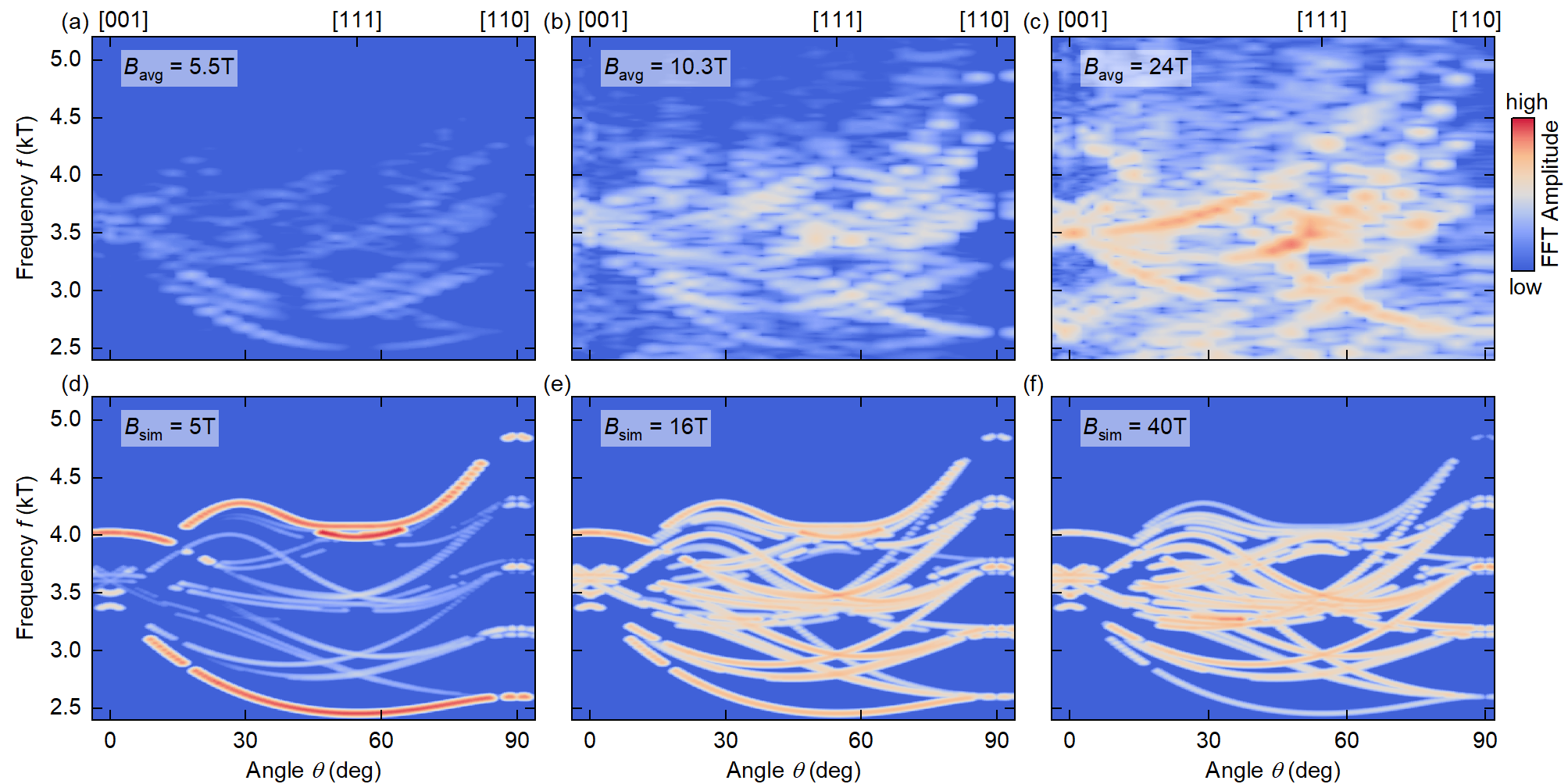}
  \caption{Field dependence of MB at the R-point. (a)-(c)~Experimental dHvA spectra evaluated for different field ranges. The mean inverse field is stated in each panel. (d)-(f)~Simulated MB branches at different magnetic fields. The small breakdown probabilities at low fields lead to the largest amplitudes for branches involving no MB. With increasing field strength the spectral weight shifts toward branches in the center of the frequency region. This behavior is generic for MB between nested FS sheets of the same charge carrier type.}
  \label{fig:MB_Comparison_Fields}
\end{figure*}

We further analyzed the frequency branches attributed to the R-point with respect to their dependence on the magnetic field range used in the analysis. Figs.~\ref{fig:MB_Comparison_Fields}(a) through \ref{fig:MB_Comparison_Fields}(c) show the dHvA spectra in the relevant frequency regime evaluated using different field windows. The size of the window in 1/B was kept constant and the average inverse field is stated in each panel. For the lowest field range, shown in Fig.~\ref{fig:MB_Comparison_Fields}(a), the low-frequency branch corresponding to FS trajectories on the inner FS pockets without MB is clearly visible. Further frequency branches corresponding to MB trajectories with comparable amplitude are visible at larger frequencies. The upper frequency branch corresponding to FS trajectories on the outer FS sheets is not visible in this field range. The decreasing amplitude for larger frequencies is explained by a stronger Dingle damping due to the larger cyclotron orbit.

For an average field of 10.3\,T, depicted in Fig.~\ref{fig:MB_Comparison_Fields}(b), frequency contributions in the whole range between the two enveloping frequency branches emerge.
For the largest fields analyzed, shown in Fig.~\ref{fig:MB_Comparison_Fields}(c), further spectral weight shifts from the enveloping frequency branches to MB trajectories as expected from the increased probability of MB at larger fields.

We performed MB calculations for different fields $B_\mathrm{sim}$. The results are shown in Figs.~\ref{fig:MB_Comparison_Fields}(d) through \ref{fig:MB_Comparison_Fields}(f). Consistent with experiment, the largest contribution at low fields is given by the two non-MB frequency branches. Upon increasing the field, spectral weight shifts towards MB branches between the non-MB branches that exhibit the largest expected oscillation amplitudes at high fields.
The field evolution of the spectral weight further confirms the identification of the observed frequency branches as MB trajectories involving partial MB.

\begin{figure}
  \includegraphics[width=\linewidth]{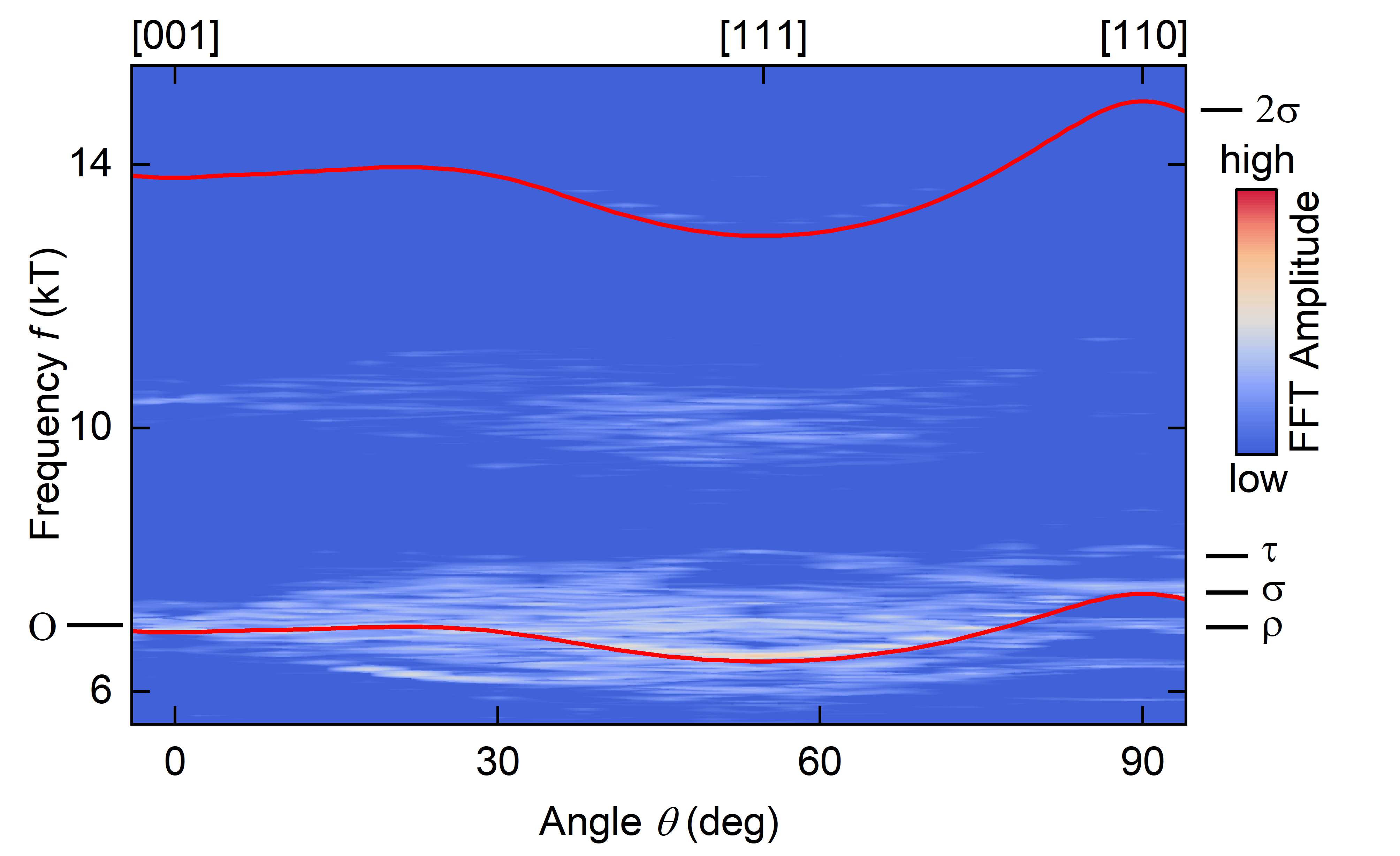}
  \caption{Higher order MB at the R-point. SdH spectra in the frequency regime containing the second, third, and fourth order MB frequencies. The sum and twice the sum of the non-MB orbits are indicated by colored lines.}
  \label{fig:SdH_sigma_2sigma}
\end{figure}

Next, we discuss the high-frequency regime involving branches arising from higher order MB orbits, i.e., trajectories that include more than one revolution around the R-point.
With the first order trajectories leading to frequencies in the range between 2.7\,kT and 4.8\,kT, the second order trajectories are expected between 5.4\,kT and 9.6\,kT. In the experimental spectra, shown in Fig.~\ref{fig:SdH_sigma_2sigma}, the second order trajectories correspond to frequency branches connecting $O$ and $\rho$ to $\tau$. Additionally, the third and fourth order MB branches are visible at larger frequencies.

We note that the origin of frequencies in these regimes differs subtly from that of conventional harmonics. In the semiclassical Onsager picture, harmonic frequencies may be though of as arising from trajectories involving more than a single revolution around the Fermi surface on the same orbit. In the case of the MB orbits centered around the R-point in PdGa, quasiparticles do not necessarily follow the same orbit for each revolution around the FS.
For instance, this can be seen by the number of lobes included in the MB trajectories. As discussed above, each lobe contributes an area corresponding to a frequency of $\approx$~0.6\,kT. Hence, frequencies $\xi$ and $o$, which enclose two and three lobes, respectively, are separated by 0.6\,kT. For second harmonics in the conventional sense one would thus expect a separation of 1.2\,kT. In contrast, the difference between $\sigma$ and $\tau$ is also close to 0.6\,kT indicating that the underlying trajectories include four and five lobes, respectively. $\tau$ is thus not a harmonic of $o$ but a new breakdown trajectory.
We observe the same phenomenology for the third order (fourth order) MB frequencies shown in Fig.~\ref{fig:SdH_low_T_FFTs}(b) with frequencies $\psi_i$ and $\sigma+\xi$ ($\omega_i$ and $2\sigma$) evenly separated by $\approx$~0.6\,kT.

As a general observation, the spectral weight in each MB region increasingly shifts towards MB branches at the center of each region with increasing order. The reason for this shift is twofold.
First, MB branches with higher orders exhibit a stronger temperature and Dingle damping than the lower orders and are thus only detected at higher magnetic fields. Consequently, the breakdown probability at every MB junction is larger, shifting more weight towards the central branches, cf. Fig.~\ref{fig:MB_Comparison_Fields}.
Second, since the number of MB junctions increases with order, the statistical weight shifts towards frequency branches with larger multiplicity, i.e., frequency branches for which multiple trajectories leading to the same frequency exist.

To illustrate this point, we examine the case of magnetic field applied along the $[110]$ direction, as depicted in Fig.~\ref{fig:MB-R-point}(b). There are four MB junctions between the inner and outer pair of trajectories. All MB orbits enclose the area defined by the inner FS sheets and may additionally encompass an area defined by the number of lobes included.
The outermost frequencies $\lambda$ and $\pi$ correspond to the case of no MB at every junction, hence their multiplicity is 1. Frequency $\nu$ encompasses one lobe and because there are four lobes in total, the multiplicity of this MB orbit is 4. The same holds true for frequency $o$ encompassing three lobes. The highest statistical weight lies on frequencies $\xi$, encompassing two lobes, for which the multiplicity is 6 (two trajectories for which the lobes are opposite and four for which they are adjacent).
For higher order MB trajectories, the difference in multiplicities is even more pronounced. While the outermost frequency branches always have a multiplicity of 1, the multiplicities of frequency branches between them increases in general. Considering the second order, the centremost frequency branch $\sigma$ encloses four out of a total of eight lobes. Hence, its multiplicity is 70 and its relative weight in the second order MB region is therefore expected to be larger than that of $\xi$ in the first order MB region.

The MB branch corresponding to the sum of the lowest and highest frequency branches is particularly pronounced for even order MB regions. For the second order, this branch connects frequencies $O$ and $\sigma$ and dominates the spectrum in this region. For the fourth order, it represents the only frequency branch clearly discernible. In the frequency regimes containing the first order and third order MB, this branch is present but exhibits an amplitude consistent with the MB simulations shown in Fig.~\ref{fig:MB-R-point}(d) without additional enhancement.

We attribute the large amplitude of this frequency branch to MB trajectories that include an uneven number of tunneling events for one revolution around the FS pockets. While for all odd orders these MB trajectories do not constitute a closed orbit and therefore do not contribute to the QO spectrum, they form a closed trajectory for an even number of revolutions around the R-point. Hence, there are additional closed MB trajectories that are allowed for the even order MB trajectories. The particular enhancement of the sum frequency branch indicates that MB for these trajectories takes place at the same MB junctions for each revolution around the R-point.

\section{Conclusions}\label{sec:Conclusion}
We performed a comprehensive study of dHvA and SdH oscillations in PdGa. The oscillation frequencies and cyclotron masses detected match the Fermi surface obtained from DFT calculations exceptionally well when taking into account (i)~the nodal plane degeneracies on the BZ boundary, (iii)~band splitting induced by spin-orbit coupling, and (iii)~partial magnetic breakdown between the individual pockets. We explicitly calculated magnetic breakdown spectra including extremal MB trajectories away from high-symmetry planes and analyzed higher order MB orbits involving more than one revolution around the Fermi surface which are distinct from conventional harmonics.
Our findings provide a complete description of the FS of PdGa comprising multiple nested and partially intersecting pockets split by spin-orbit coupling and highlight the relevance of nodal planes and magnetic breakdown in the interpretation of quantum oscillation spectra.

\begin{acknowledgments}
	We gratefully acknowledge the support of the LNCMI-CNRS, member of the European Magnetic Field Laboratory (EMFL) as well as financial support of DFG via TRR360 - 492547816 (ConQuMat), SPP 2137 (Skyrmionics) under grant no. PF393/19 (project id 403191981), DFG-GACR project WI3320/3-1 (project id 323760292), ERC Advanced grant no 788031 (ExQuiSid) and Germany's excellence strategy EXC-2111 390814868. I.V. acknowledges support by the DAAD through the Stipendienprogramm Deutsche Auslandsschulen.
\end{acknowledgments}

\end{document}